\pdfoutput=1
\documentclass[aps,prb, twocolumn, superscriptaddress,nofootinbib]{revtex4-2}

\usepackage{graphicx} 
\usepackage{amsthm} 
\usepackage{amsmath,braket}
\usepackage{mathtools}
\usepackage{bbold}  
\usepackage{amsfonts} 
\usepackage{graphicx}
\usepackage{hyperref}
\usepackage[dvipsnames]{xcolor} 
\usepackage{tikz}
\usepackage{tikz-cd} 
\usetikzlibrary{arrows.meta}
\usetikzlibrary{positioning}

\usepackage{adjustbox} 
\usepackage{centernot}
\usepackage[english]{babel}
\usepackage[T1]{fontenc}
\usepackage[normalem]{ulem} 
\usepackage[%
labelfont=bf,            
  textfont=small,     
  justification=justified,
  singlelinecheck=false    
]{caption}
\DeclareCaptionJustification{justified}{\leftskip=0pt \rightskip=0pt \parfillskip=0pt plus 1fil}
\usepackage{subcaption}
\usetikzlibrary{decorations.pathreplacing}
\usetikzlibrary{decorations.pathmorphing}

\theoremstyle{definition}

\begin{document}
 
\title{Measurement-Induced Phase Transition in a Disordered XX Spin Chain: A Real-Space Renormalization Group Study}

\author{Siddharth Tiwary}
\email{siddharthtiwary@berkeley.edu}
\affiliation{Department of Physics, 366 Physics North MC 7300, University of California, Berkeley, CA 94720, USA}
\author{Joel E. Moore}
\email{jemoore@berkeley.edu}
\affiliation{Department of Physics, 366 Physics North MC 7300, University of California, Berkeley, CA 94720, USA}
\affiliation{Materials Sciences Division, Lawrence Berkeley National Laboratory, Berkeley, CA 94720, USA}

\begin{abstract}
Spin chains with quenched disorder exhibit rich critical behavior, often captured by real-space renormalization group (RSRG) techniques. However, the physics of such systems in the presence of random measurements (i.e., non-Hermitian dephasing) remains largely unexplored. The interplay between measurements and unitary dynamics gives rise to novel phases and phase transitions in monitored quantum systems. In this work, we investigate the disordered XX spin chain subject to stochastic local measurements in the $X$ and $Y$ bases. By mapping the monitored chain to a non-Hermitian spin ladder with complex couplings, we propose an RSRG-for-excited-states (RSRG-X) approach for this open-system setting. Our analysis reveals a new class of strongly disordered fixed points that emerge due to non-unitarity, broadening the landscape of critical phenomena accessible via RSRG.

\end{abstract}
                
\maketitle

\section{Introduction}
Systems with quenched disorder have been analyzed in the past using a variety of techniques, including the replica trick~\cite{replica1,replica2,replica3}, supersymmetry~\cite{susy1,susy2,susy3}, and the real space renormalization group~\cite{rsrgMain,rsrgOrigin,rsrgReviewOld,rsrgReviewNew} (RSRG). The RSRG approach put forward by Ma and Dasgupta~\cite{rsrgOrigin} involves decimating out the strongest random coupling in a given disordered system by projecting the system to a subspace of interest in the dominant coupling's eigenspace and treating the rest of the system perturbatively.

As this process is carried out iteratively, if the distribution of couplings broadens without limit, the perturbation theory ends up being asymptotically exact as the decimated dominant couplings end up much larger than the nearby couplings in the system, justifying the perturbative approach. In this limit,  RSRG offers insight into the universal properties of the random system, like infinite disorder fixed points (IDFPs) and Griffiths phases~\cite{rsrgReviewNew,rsrgReviewOld,Griffiths2,Griffiths3}. 

While the original RSRG approach was used to study ground state properties by projecting the strongest bond into its ground state at each decimation step, the RSRG for excited states (RSRG-X) constructs the spectrum of the Hamiltonian by being flexible at each decimation step about which eigenstate of the strongest bond is retained~\cite{rsrgX1,rsrgX2}. The choice of eigenstate at each step creates a branched `decision tree,' with each path down the tree leading to a different many-body eigenstate in the spectrum. In this way, RSRG-X provides a systematic procedure for building excited eigenstates of strongly disordered systems, enabling the study of dynamical properties, entanglement structure, and phase transitions beyond the ground state. Effectively, the energy scale integrated out as renormalization proceeds is the largest local energy gap between the highest and lowest energy states, and we choose whether to project onto the highest or lowest excited state st each decimation step.

More recently, the RSRG approach has been applied to deal with many-body localization, dynamical phase transitions, and quantum information spreading~\cite{rsrgAdd5,rsrgAdd6,rsrgAdd7,rsrgAdd8,rsrgAdd9}. First created to deal with Hamiltonian problems~\cite{rsrgAdd1,rsrgAdd2,rsrgAdd3,rsrgAdd4,rsrgMain,rsrgOrigin}, RSRG has been extended to deal with driven~\cite{rsrgDriven,monthus2} and dissipative~\cite{rsrgdiss1,rsrgdiss2,rsrgdiss3,rsrgdiss4,rsrgdiss5,rsrgdiss6} systems. The effects of dissipation on quantum criticality have been investigated using RSRG on disordered Ising spin chains coupled to oscillator baths~\cite{rsrgdiss1,rsrgdiss2,rsrgdiss3,rsrgdiss4} and on the discretized Landau-Ginzburg free energy functional in real space~\cite{rsrgdiss5,rsrgdiss6}. While ordered spin chains have been studied extensively using the Lindblad formalism~\cite{lindblad1,lindblad2,lindblad3,lindblad4}, disordered systems with bulk Lindblad dephasing~\cite{monthus1} do not appear to have been studied using the RSRG technique.

The introduction of random measurements tends to destroy entanglement in many-body systems and creates competition between dephasing and unitary evolution that leads to exotic phases in monitored systems. Measurement-induced phase transitions (MIPTs) in such systems can often be detected using the entanglement entropy, which transitions from a volume-law to an area-law phase if measurements are frequent enough.~\cite{mipt1,mipt2,mipt3,mipt4} The experimental observation of these MIPTs, however, has been challenging due to the need to post-select based on the measurement outcome, with the probability of the target outcome shrinking exponentially with the number of measurements.

We study a disordered XX spin chain with random single-site $X$ and $Y$ measurements, with the measurement probabilities themselves picked from a random distribution: similar to Zabalo \textit{et al}~\cite{rucMIPT}, who studied a random unitary circuit using RSRG in a setup where tuning the probability distribution of measurement probabilities led to an MIPT flanked by strongly disordered Griffiths phases. The results of measurements are discarded. In the continuum (long time) limit, the setup can alternatively be described as a disordered $XX$ chain with randomly picked $X$- and $Y$- dephasing rates at every site.

Our model maps to a disordered non-Hermitian XX spin ladder~\cite{monthus1,monthus2,spinLadders,ladderMap}. We propose an RSRG-X scheme that could be employed to study a portion of the spectrum containing an exponentially large number of states. We focus our attention on one particular fast-oscillating, long-lived mode of the spin ladder, analyzing the flow of distributions of couplings to find strong disorder fixed points with long-range correlations and area law entanglement that are not observed in the non-measured XX model, and transitions between them driven by the rate of measurements. (Studying the longest-lived states is commensurate with coarse-graining the random measurements over time to take the continuum limit.) Since our setup discards the measurement outcomes anyway, the fixed points we predict would not suffer from the post selection problem if implemented. In effect, our technique also works for a reflection-symmetric (Hermitian) XX spin ladder, and the fixed points we find translate directly to RSRG fixed points of the ground state of such a ladder.   

RSRG provides an asymptotically exact framework for analyzing strongly disordered systems, making it well suited to capture universal features in the presence of both disorder and dissipation. Studying the monitored XX chain under strong disorder conditions reveals novel phases that are absent in closed or homogeneous dissipative systems. Furthermore, the exact diagonalization of $k$-local Hamiltonians is QMA-complete for $k\geq2$~\cite{qma1,qma2,qma3}. Even when the locality is further restricted to a one-dimensional chain and one asks only about the ground-state, the local-Hamiltonian problem remains QMA-complete, as proved by Gottesman and Irani~\cite{qma4}. This makes the numerical study of disordered systems generically difficult, even on a quantum computer. Analytical insight from techniques like RSRG into the physics of a class of disordered quantum systems is hence valuable.

The rest of the paper is organized as follows: in section \ref{setup}, we describe the model and its mapping to a spin ladder with complex couplings. In section \ref{rules}, we look at the decimation rules for the spin ladder, finally followed by an analysis of a set of fixed points of the system in section \ref{fp}. We conclude with a discussion of other avenues, namely free fermionic spin chains and the reflection-symmetric XX ladder, where our analysis could help uncover the physics. We briefly discuss platforms that could be used to realize the disordered monitored XX spin chain.

\section{Setup}
\label{setup}
We consider an $N$-site XX spin-$1/2$ chain with periodic boundary conditions, $\mathcal H=\sum_{i=1}^N -J_i (X_i X_{i+1}+Y_iY_{i+1})$, where $X_i$ and $Y_i$ are Pauli $X$ and $Y$ operators at the $i^\text{th}$ site, and the couplings $J_i$ are drawn from a random distribution. The operators $X_i$ and $Y_i$ are measured with probability $p_i$ per unit time, and the measurement results are discarded. The probabilities $\{p_i\}$ are themselves picked from a random distribution. For simplicity, we restrict ourselves to even $N$. Then the Hamiltonian anticommutes with $\prod_{i=1}^{N/2}Z_{2i}$, and eigenvalues come in pairs with opposing signs: the spectrum is symmetric about zero energy.

Let $\mathbb P_i^{x,+}$ and $\mathbb P_i^{x,-}$ be projectors onto the eigenspaces of $X_i$, and $\mathbb P_i^{y,+}$ and $\mathbb P_i^{y,-}$ onto those of $Y_i$. Over a time period $\delta t\ll \min(1/(\sum_i p_i), 1/E_\text{max})$, where $E_\text{max}$ is the largest eigenvalue of the Hamiltonian, the evolution of the state of the system $\rho(t)$ under unitary evolution and random measurements can be notated,

\begin{multline}
    \rho+\frac{d\rho}{dt}\delta t=(1-2\sum_i p_i\delta t)(1-i H \delta t)\rho(1+iH\delta t)\\+\delta t \sum_i p_i [(P_i^{x,+}\rho P_i^{x,+}+P_i^{x,-}\rho P_i^{x,-}-\rho)\\+(P_i^{y,+}\rho P_i^{y,+}+P_i^{y,-}\rho P_i^{y,-}-\rho)].
\end{multline}

In the continuum limit over a coarse-graining time period $t\gg 1/p_i$, this yields for the evolution of the state $\rho$ the Lindblad equation,

\begin{equation}
    \frac{d\rho}{dt}=-i[\mathcal H,\rho]+\sum_{i=1}^N\frac{p_i}{2}\left(\mathcal D[X_i]\rho+\mathcal D[Y_i]\rho\right),\label{eom}
\end{equation}
with $\mathcal D[A]\rho\equiv A\rho A^\dag-\frac{1}{2}\{A^\dag A,\rho\}$, signifying dephasing along the measured operators at the measurement rate.

We proceed to vectorize this equation. Write out the matrix elements of $\rho$ in the Pauli-Z basis, 

\begin{multline}\rho(t)=\sum_{S_1=\pm 1/2,\dots ,S_N=\pm 1/2}\;\sum_{T_1=\pm 1/2,\dots ,T_N=\pm 1/2}\\\rho_{S_1\dots S_N T_1\dots T_N}\ket{S_1\dots S_N}\bra{T_1\dots T_N}.\end{multline}
Thus define the vectorized form of $\rho$,
\begin{multline}\ket{\rho(t)}\equiv\sum_{S_1=\pm 1/2}\dots\sum_{S_N=\pm 1/2}\sum_{T_1=\pm 1/2}\dots\sum_{T_N=\pm 1/2}\\\rho_{S_1\dots S_N T_1\dots T_N}\ket{S_1\dots S_N}\ket{T_1\dots T_N}.\end{multline}
Clearly, a general term $A\rho B$ for some operators $A$ and $B$ maps to $A\otimes B^T \ket{\rho}$, with the operator $B$ on the right getting transposed. In Liouville space, Eq.\eqref{eom} then becomes
\begin{multline}
    \frac{d\ket{\rho}}{dt}=\mathcal L\ket{\rho}\equiv [-i(\mathcal H\otimes \mathbb I-\mathbb I\otimes \mathcal H^T)+\\\sum_{i=1}^N\frac{p_i}{2}\left(X_i\otimes X_i-Y_i\otimes Y_i-2\mathbb I\right)]\ket{\rho},\label{vec}
\end{multline}

where $\mathbb I$ is the identity operator and we have defined the generator of time translations in the superoperator formalism, $\mathcal L$. 

\begin{figure}
    \centering
    \includegraphics[width=\linewidth]{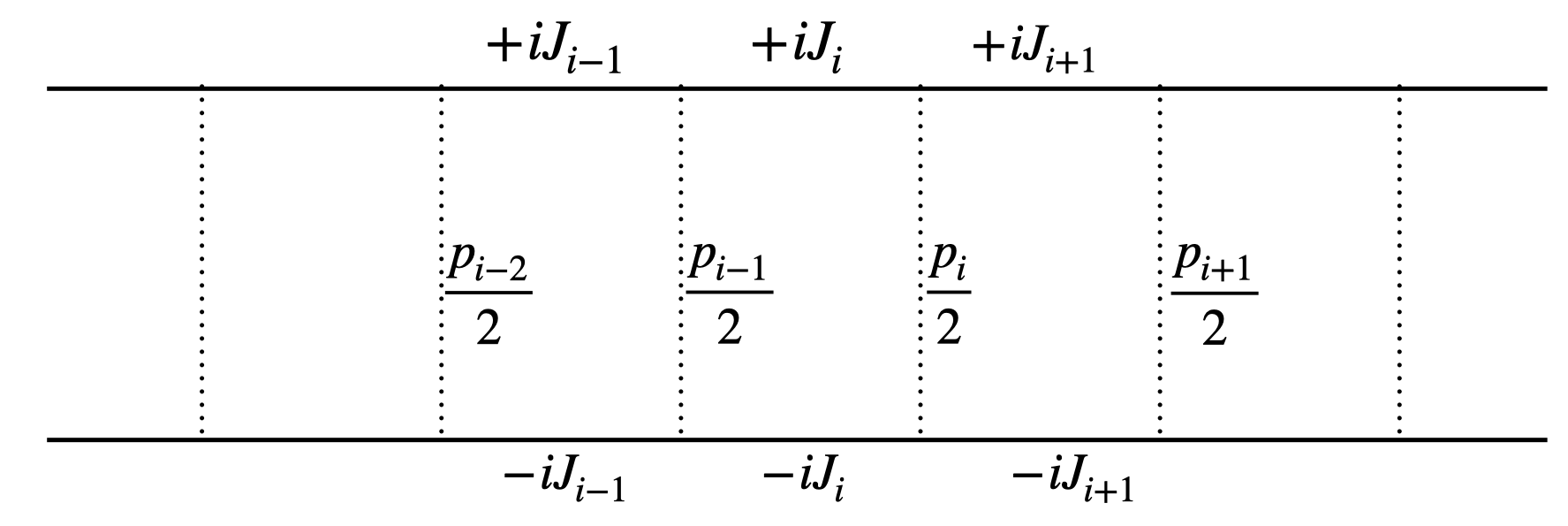}
    \caption{Mapping the dissipative XX spin chain to a non-Hermitian spin ladder: Horizontal bonds (imaginary) represent Hermitian XX interactions with coupling strengths $J_i$. Vertical bonds correspond to dephasing rates $p_i/2$.}
    \label{Fig1}
\end{figure}

Let's flip the qubits $\ket{0}\leftrightarrow\ket{1}$ in the Hilbert space corresponding to the bras (i.e. the right side of the tensor products above). This process takes $\mathbb I\to\mathbb I, \,X_i\to X_i,\,Y_i\to -Y_i$, and $Z_i\to -Z_i$ in the second Hilbert space. Under this transformation, the original density matrix then becomes $\ket{\rho\sum_i X_i}\equiv \ket{\rho}\rangle$. In this form, Eq.\eqref{vec} can be interpreted as a non-Hermitian XX Hamiltonian (the Liouvillian) for a spin ladder, with Hamiltonian (purely imaginary in our convention) couplings $\{\pm iJ_i\}$ along horizontal and dissipative (real) couplings $\{p_i/2\}$ along vertical bonds; see Fig.(\ref{Fig1}). For notational convenience, we denote the $X$s, $Y$s, and $Z$s in the upper rail of the ladder as $\mathcal X$, $\mathcal Y$, and $\mathcal Z$, and those in the lower rail as $x$, $y$, and $z$. Thus, write the Liouvillian,
\begin{multline}
   \mathcal L= \sum_{i=1}^N iJ_i \left[(\mathcal X_i \mathcal X_{i+1}+\mathcal Y_i \mathcal Y_{i+1})-(x_ix_{i+1}+y_iy_{i+1})\right]
  \\+\sum_{i=1}^N\frac{p_i}{2}\left(\mathcal X_i x_i+\mathcal Y_i y_i-2\mathbb I\right).\label{vec2}
\end{multline}

We may diagonalize to find a spectral decomposition in terms of right and left eigenvectors, $\mathcal L=\sum_n \lambda_n \ket{\lambda_n^R}\bra{\lambda_n^L}$, where $\mathcal L\ket{\lambda_n^R}=\lambda_n\ket{\lambda_n^R}$ and $\bra{\lambda_n^L}\mathcal L=\lambda_n\bra{\lambda_n^L}$. The eigenvectors are orthonormal: $\braket{\lambda_m^L|\lambda_n^R} = \delta_{mn}$ and provide a resolution of the identity: $\mathbb I=\sum_n\ket{\lambda_n^R}\bra{\lambda_n^L}$. The eigenvalues $\lambda_n$ are complex, with real parts corresponding to decay rates and imaginary parts to oscillation frequencies. The state we select in our RSRG algorithm will, without loss of generality, seek to maximize the coherent phase oscillation frequency and maximize the lifetime of the state at each RG step. We should thus expect it to take us to some $\lambda_n$ in the bottom right corner of the complex plane: a relatively fast-oscillating, relatively long lived mode, since in general there will not exist a mode that simultaneously maximizes both oscillation frequency and lifetime. These choices give us a mode that is analytically tractable to work with. Physically, we also expect modes with high oscillation frequencies to be more robust against slow, external perturbations, as rapid oscillations tend to average out low-frequency noise.

Measurements break the strong $U(1)$ symmetry generated by the total spin, $\sum_{i=1}^N Z_i$, to a weak $U(1)$ symmetry generated by $\sum_{i=1}^N (Z_i+z_i)$~\cite{strongweak}, and likewise for parity, generated by the product of all spin operators. The maximally mixed state is a steady state solution to Eq.\eqref{eom}, which implies that the corresponding Liouville space vector $\ket{\mathbb I}\rangle$ is an eigenstate of $\mathcal L$ with vanishing eigenvalue. Likewise, the parity operator, $\prod_{i=1}^N Z_i$ (or $\ket{\prod_{i=1}^N Z_i}\rangle$ in Liouville space) is an eigenstate of $\mathcal L$ with eigenvalue $-2\sum_{i=1}^N p_i$, the fastest decaying mode of the system (and a purely decaying, i.e. zero-energy mode at that). Additionally, $\ket{\prod_{i=1}^N X_i}\rangle$ and $\ket{\prod_{i=1}^N Y_i}\rangle$ are purely decaying modes with eigenvalue $-\sum_{i=1}^N p_i$. Except for the offset due to the identity matrix, the Lindbladian anticommutes with $\prod_{i=1}^{N/2}z_{2i-1}Z_{2i}$, leading to symmetry in the spectrum.

In the original Hilbert space we may expand the time evolution of the density matrix into eigenmodes of the Lindbladian, 
\begin{equation}
    \rho(t)=\sum_kc_k\rho_{SS}^{(k)}+\sum_ld_l\rho^{(l)}e^{i\Im\lambda_l t-|\Re\lambda_l t|},
\end{equation}
where $\rho_{SS}^{(k)}$ furnish a basis spanning the space of steady states of the density matrix with $c_k$ appropriately normalized, and $\rho^{(l)}$s are traceless oscillatory modes that can decay and vanish, with frequencies and decay rates governed by the imaginary and real parts of their eigenvalues. Our approach will in principle enable the study of between $2^{(N/2)}$ and $2^N$ of these modes, but we will focus on modes with large negative frequencies and small decay rates. Within such a mode, spins shall see (possibly long range) correlations that might yield to analysis via RSRG and new kinds of quantum criticality.

\section{RSRG Decimation Rules}
\label{rules}
At each step, we pick the largest bond in the system, $\max\{J_i,p_i/2\}$, and treat the neighboring couplings as perturbations. We project the subsystem coupled by this largest bond to its ground (most negative eigenvalue) state, if the coupling is imaginary, or to the longest-lived state, if the coupling is real, and consider the couplings generated by second-order perturbation theory in this subspace. This choice will take us to a long-lived, fast-oscillating mode in the spectrum.

We iterate the process of decimating out the largest bond and study how the couplings flow under this RG flow. The process is asymptotically exact if it takes us to an infinite disorder fixed point (IDFP), as we discuss in the next section, i.e. at an IDFP, RSRG gets more and more accurate as the width of the distribution of couplings increases so the largest coupling is always much larger than its neighbors and perturbation theory is exact.

\begin{figure}
    \centering
    \includegraphics[width=\linewidth]{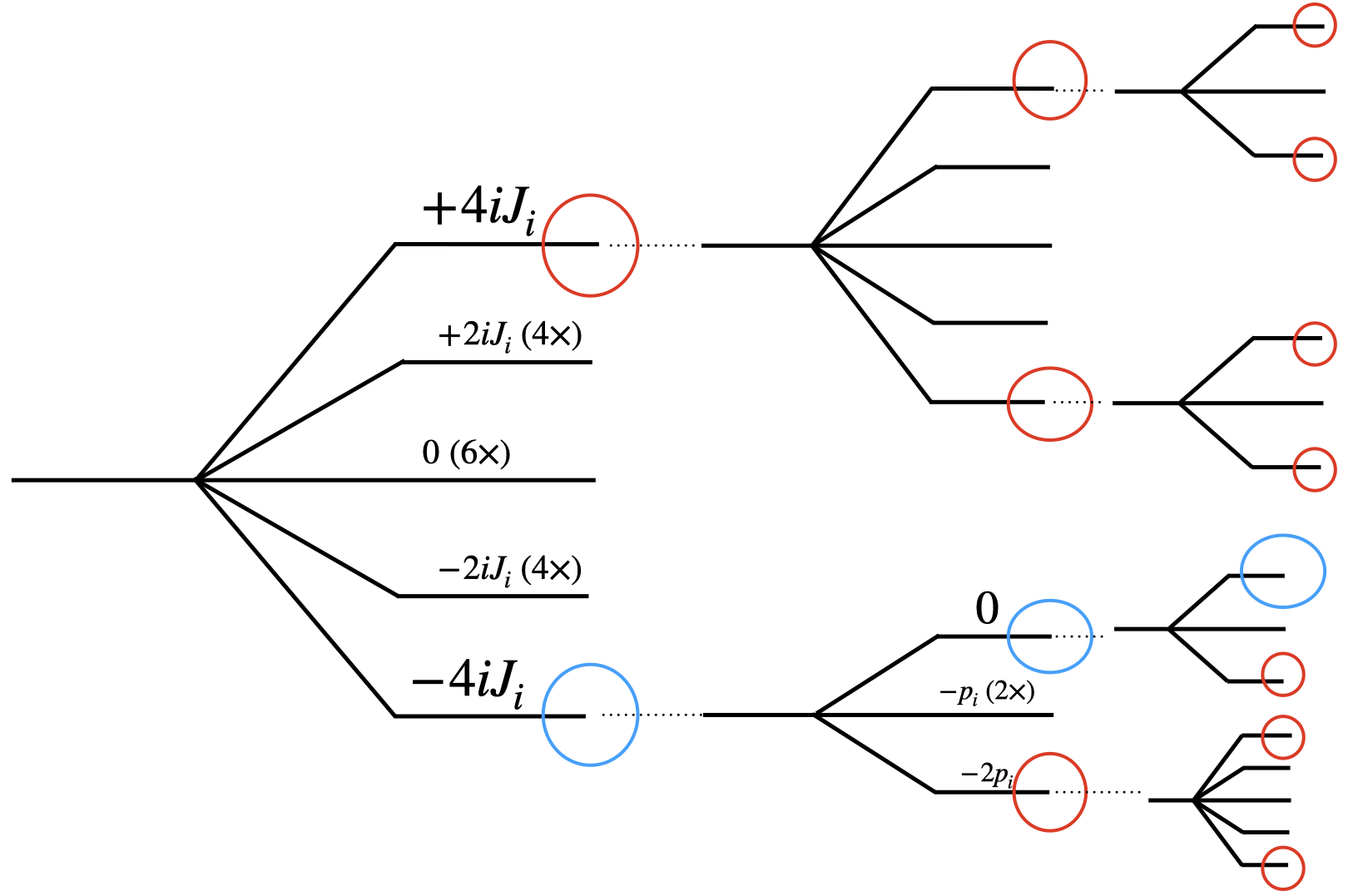}
    \caption{Schematic depiction of the spectrum as we perform RSRG-X on the spin ladder: at each step, eigenstates of the strongest bond with the maximum and minimum absolute eigenvalues (circled) are nondegenerate, and constitute possible routes our algorithm could take us through. The decimation of $J_i$s produces 5 levels (a few example eigenvalues and their degeneracies being listed), while the decimation of $p_i$s produced 3; The ones circled blue signify the sequence we take in this paper as we minimize the imaginary eigenvalues (i.e. negative coherence frequencies) and maximize the real (lifetime) ones at every step.}
    \label{Fig1.5}
\end{figure}

The algorithm presented would go through even if we picked the shortest-lived or the most excited state at each decimation step, since it only relies on the non-degeneracy of the state we project into. Therefore, at each decimation step, we really have two actions to choose from: the state with the highest, or the lowest eigenvalue. Therefore, this RSRG-X method could be applied to study exponentially many states in the spectrum.  This arbitrariness is illustrated by Fig.(\ref{Fig1.5}), which shows the possible routes RSRG-X can take us through, and the particular route we pick as we decimate couplings.

More explicitly, separate $\mathcal L=\mathcal L_0+\mathcal L_\text{pert}$. When the strongest coupling in the system is $J_i$, \begin{equation}\mathcal L_0=iJ_i \left[(\mathcal X_i \mathcal X_{i+1}+\mathcal Y_i \mathcal Y_{i+1})-(x_ix_{i+1}+y_iy_{i+1})\right].\end{equation} Meanwhile, 
\begin{multline}\mathcal L_\text{pert}=\\iJ_{i+1}[(\mathcal X_{i+1} \mathcal X_{i+2}+\mathcal Y_{i+1} \mathcal Y_{i+2})-(x_{i+1}x_{i+2}+y_{i+1}y_{i+2})]\\+iJ_{i-1}[(\mathcal X_{i-1} \mathcal X_{i}+\mathcal Y_{i-1} \mathcal Y_{i})-(x_{i-1}x_{i}+y_{i-1}y_{i})]\\+\frac{p_i}{2}\left(\mathcal X_i x_i+\mathcal Y_i y_i-2\mathbb I\right)+\frac{p_{i+1}}{2}\left(\mathcal X_{i+1} x_{i+1}+\mathcal Y_{i+1} y_{i+1}-2\mathbb I\right)\\+\dots,\end{multline}
where the dots represent terms that act trivially on the eigenbasis of $\mathcal L_0$, while the terms we explicitly wrote out are instrumental in generating the new couplings.

$\mathcal L_0$ represents the imaginary part of the ladder Hamiltonian, and as such has purely imaginary eigenvalues, so we seek to project to its ground state to obtain the most negative eigenvalue. Its ground state is the tensor product of a singlet and a triplet; denote it as $\ket{\text{GND}}$. Let $\ket{m}$ label a general eigenstate.

Post decimation, the updated Lindbladian in second order perturbation theory becomes, \begin{multline}
    \mathcal L'=E_\text{GND}+\bra{\text{GND}}\mathcal L_\text{pert}\ket{\text{GND}}\\+\sum_{m\neq \text{GND}}\frac{\bra{\text{GND}}\mathcal L_\text{pert}\ket{m}\bra{m}\mathcal L_\text{pert}\ket{\text{GND}}}{E_\text{GND}-E_m},
\end{multline}
where $E_\text{GND}=\bra{\text{GND}}\mathcal L_0\ket{\text{GND}}$ and $E_\text{m}=\bra{m}\mathcal L_0\ket{m}$ are eigenvalues of the unperturbed Lindbladian. The second order piece of this expression contains terms which carry information about the new coupling. The other terms serve to renormalize the net oscillation frequency and dissipation rate. When decimating $J_i$ in particular,
\begin{multline}
    \mathcal L'=\\i\frac{J_{i-1}J_{i+1}}{J_i}(\mathcal X_{i-1}\mathcal X_{i+2}+\mathcal Y_{i-1}\mathcal Y_{i+2}-x_{i-1}x_{i+2}-y_{i-1}y_{i+2})\\+\frac{p_{i-1}}{2}(\mathcal X_{i-1}x_{i-1}+\mathcal Y_{i-1}y_{i-1})+\frac{p_{i+2}}{2}(\mathcal X_{i+2}x_{i+2}+\mathcal Y_{i+1}y_{i+1})\\+\dots,
\end{multline}
where the $\dots$ represent renormalization of oscillation frequency, along with the Lindbladian for the rest of the chain. Therefore, decimating a $J_i$ simply causes the bond strengths $\{J_{i-1}, J_i, J_{i+1}\}$ to get replaced by a single $J_{i-1}J_{i+1}/J_i$, with no new forms of couplings (e.g., diagonal) generated.

\begin{figure}
    \centering
    \includegraphics[width=0.75\linewidth]{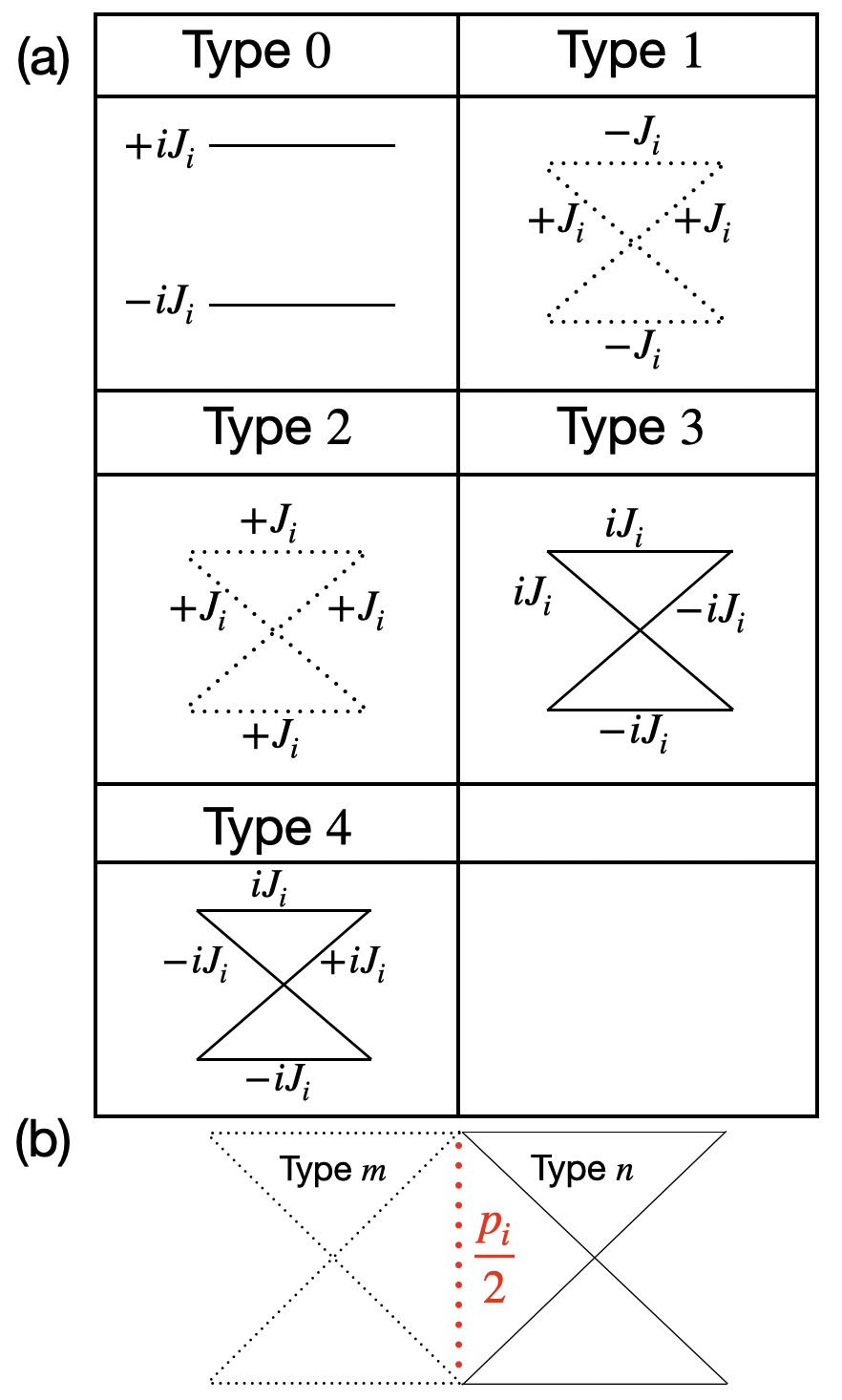}
    \caption{The structural components of the renormalized spin ladder. (a) The five distinct unit-cell configurations (Types 0–4) that emerge during the RSRG process. The system initializes with Type-0 cells (purely horizontal couplings); decimating vertical bonds generates the diagonal couplings characterizing Types 1–4. (b) The vertical dephasing bond (strength $p_i/2$, shown in red) that serves as the link connecting adjacent unit cells. The full ladder is constructed by chaining the unit cells from (a) together via the vertical bonds shown in (b).}
    \label{Fig2}
\end{figure}
\begin{figure}
    \centering
    \includegraphics[width=\linewidth]{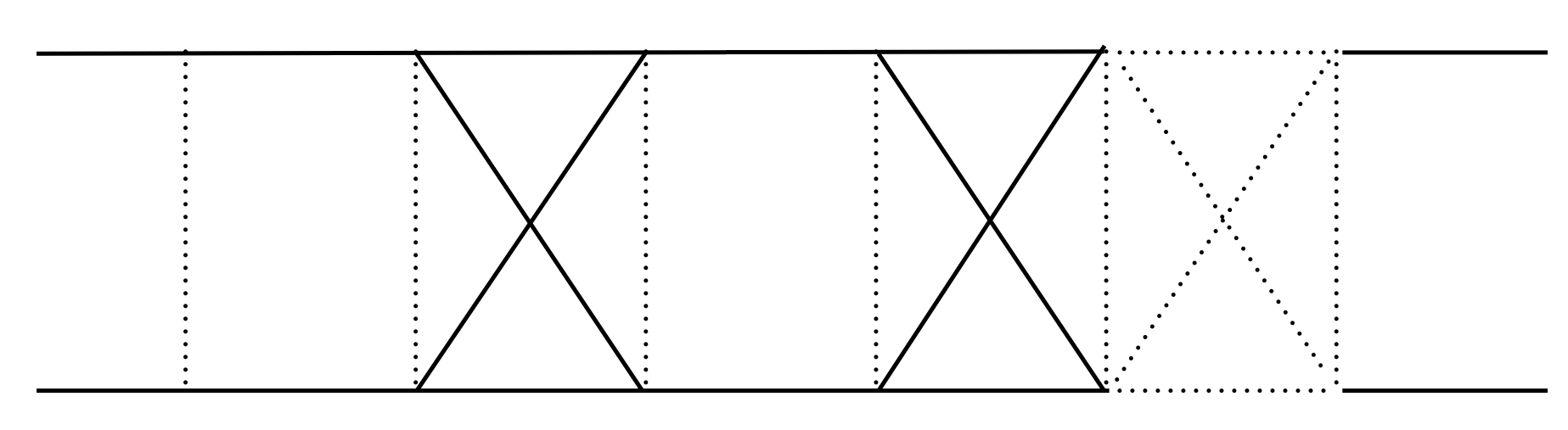}
    \caption{Example configuration of the non-Hermitian spin ladder during an intermediate stage of the RSRG process. The ladder now contains various types of cells (type-0 through type-4), each distinguished by the diagonal and horizontal couplings. Dotted lines indicate imaginary (Hermitian) couplings while solid lines stand for real (dissipative) couplings.}
    \label{Fig3}
\end{figure}

We may derive similar rules for when the vertical bond, $p_i/2$ (see Fig.(\ref{Fig2})(b)) is the largest coupling in the system and consequently decimated. Selecting the longest lived state for a single vertical bond translates to projecting the corresponding site in the spin chain onto a maximally mixed state. 

\begin{figure}
    \centering
    \includegraphics[width=\linewidth]{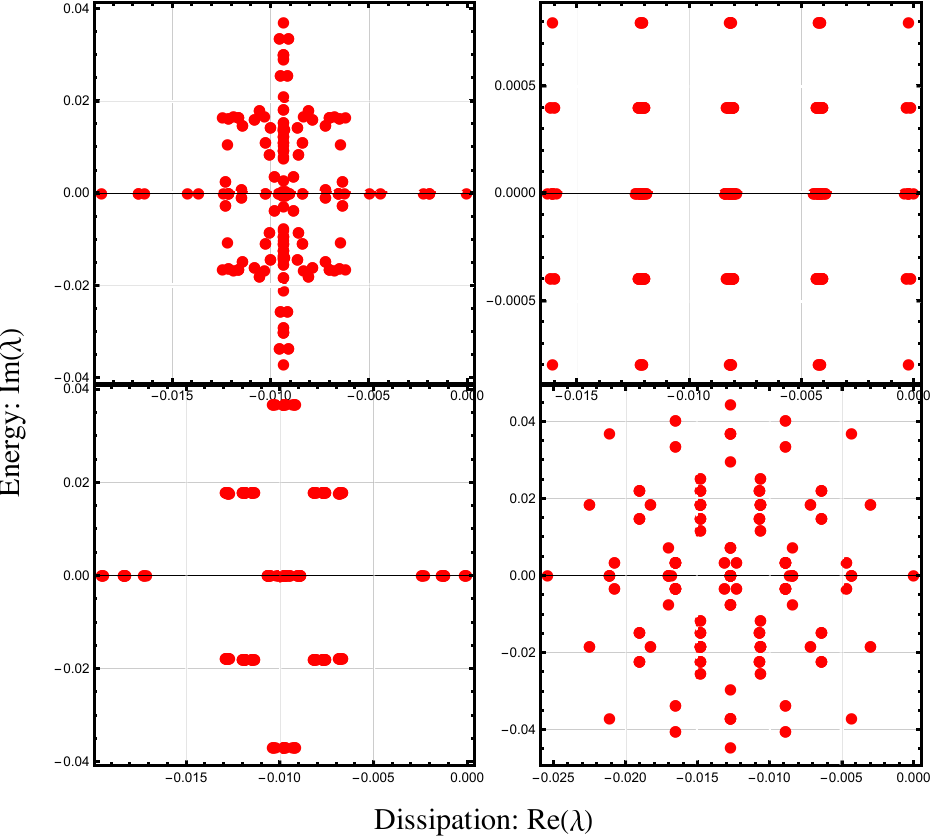}
    \caption{Numerically computed spectrum of the Lindbladian for $N=4$ rungs of the ladder, shown for 4 different random sets of couplings. $J_i$ and $p_i$ are both drawn from a strong-disorder power-law distribution, $P(x) \propto x^{-\alpha}$, defined on the interval $[0, \Omega]$, with a cutoff $\Omega = 10^{-20}$ and exponent $\alpha = 1 + 1/\ln \Omega \approx 0.98$.} The steady (maximally mixed) state is the right-most point on the $x$-axis while the fastest oscillating state is the bottom-most point on the $y$-axis. Picking the longest lived and fastest (negative frequency) oscillating states during decimation corresponds to reaching one of the modes near the lower right corners of the plots as opposed to the rightmost or the bottom-most points, since it's not possible to simultaneously maximize both lifetime and oscillation frequency.
    \label{Fig3.5}
\end{figure}

Fig.(\ref{Fig3.5}) shows the spectrum of the Lindbladian computed numerically via exact diagonalization for $N=4$ sites on the complex plane, over several runs, with the imaginary axis corresponding to oscillation frequency and the real axis to decay. The bond strengths were picked from a broad random distribution detailed in the caption. The discussed RSRG-X approach could take us to between $4$ and $16$ of the $256$ total states of the system. Our specialization to large negative oscillation frequencies and long lifetimes would restrict us to one of the states in the bottom right corner of the diagram-- for sufficiently broad initial distributions, the state with maximal Manhattan distance from the center of the spectrum. (Note that maximizing solely the lifetime would take us to the rightmost state on the $x$-axis, while maximizing solely the (negative) oscillation frequency would take us to the bottom-most state on the $y$-axis. It is not generally possible to simultaneously maximize both. We target the manifold of relatively long-lived, relatively fast-oscillating modes and construct a single state by locally maximizing the lifetime and oscillation frequency at each decimation step. For sufficiently broad initial distributions, successive decimated couplings would differ by large amounts-- so maximizing the L1 distance from the center of the spectrum at each step ought to get us the mode with that's maximally far from the center. Such a mode will either be a fast oscillating mode in the neighborhood of the longest lived mode or a long lived mode in the neighborhood of the fastest oscillating mode, depending upon whether the strongest coupling in the system is a $p_i$ or a $J_i$.)

Observe Fig.(\ref{Fig2})(a). Initially, the ladder is composed of `unit cells' of type $0$ with horizontal bonds of magnitude $\{J_i\}$, and vertical bonds with magnitudes $\{p_i/2\}$, as shown in Fig.(\ref{Fig1}). At each step, we end up decimating either a $J_i$ or a $p_i/2$. Decimation of $\{J_i\}$ preserves the initial `all-zeroes' structure of the ladder, as we just showed. However, as detailed in the Appendix, the decimation of vertical couplings can generate diagonal couplings in the system. This leads to $4$ additional types of unit cells with non-vanishing diagonal bonds, as listed in the figure, before the decimation rules close on themselves (provided $J_i$ and $p_i$ remain real, or if generated imaginary parts are irrelevant), and no new kinds of cells can be produced via further decimation. At a certain stage of RSRG, the ladder may, for instance, look like Fig.(\ref{Fig3}), with various types of unit cells contributing. Within each cell that is not of type $0$, the magnitude of each of the diagonal and horizontal bonds is always the same. Therefore, a cell can be uniquely labeled using its type and $J_i$. The ladder is a chain of cells with vertical bonds that need to be specified.

A few example $J_i$-decimation rules are listed in Table (\ref{tab1}). Decimation of a cell with coupling strength $J_i$ involves replacing the cell, along with its left and right neighbors, with a new cell carrying a new coupling, as listed in the table. Moreover, sometimes, the left and right vertical bonds of this new cell can also end up getting renormalized. The last two columns of the table show this effect. The adjoining figure illustrates our labeling.

\begin{center}
\begin{table*}

 \includegraphics[width=0.7\textwidth]{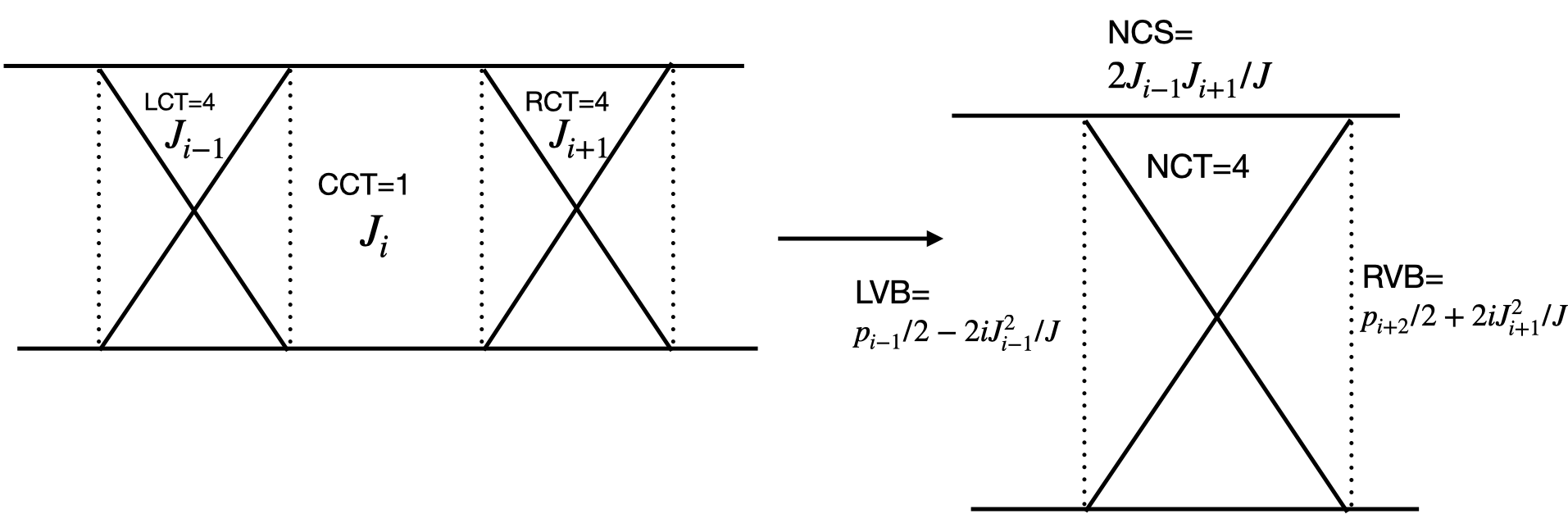}
\begin{tabular}{ |c|c|c|c|c|c|c|} 
 \hline
 \textbf{LCT} & \textbf{CCT} & \textbf{RCT} & \textbf{NCT} & \textbf{NCS} & \textbf{LVB} & \textbf{RVB} \\ 
 \hline
 0 & 0 & 0 & 0 & $J_{i-1}J_{i+1}/J_i$ & $p_{i-2}/2$ & $p_{i+1}/2$\\ 
 1 & 1 & 1  & 1 & $7J_{i-1}J_{i+1}/(2J_i)$ & $p_{i-2}/2-5J_{i-1}^2/(\sqrt{2}J_i)$ &  $ p_{i+1}/2-5J_{i+1}^2/(\sqrt{2}J_i)$\\ 
 2 & 2 & 2 & 2 & $7J_{i-1}J_{i+1}/(2J_i)$ & $p_{i-2}/2+5J_{i-1}^2/(\sqrt{2}J_i)$ & $ p_{i+1}/2+5J_{i+1}^2/(\sqrt{2}J_i)$\\ 
2 & 0 & 1 & 4 & $2J_{i-1}J_{i+1}/J_i$ & $p_{i-1}/2-2iJ_{i-1}^2/J_i$& $p_{i+1}/2+2iJ_{i+1}^2/J_i$\\
0 & 1 & 2 & -1 & $0$ & $p_{i-1}/2-9J_{i-1}^2/(8\sqrt 2 J_i)$ & $p_{i+1}/2+J_{i+1}^2/(2\sqrt 2 J_i)$\\
 \hline
 
\end{tabular}
\caption{A few rules for Decimating $J_i$; Abbreviations: LCT=Left Cell Type, CCT=Center Cell Type, RCT= Right Cell Type, NCT= New Cell Type, NCS= New Coupling Strength, LVB=Left Vertical Bond, RVB=Right Vertical Bond. The  image demonstrates the decimation scheme: the 3 cells surrounding the strongest coupling coalesce to form a single new cell with an updated type and coupling strength. A cell type of $-1$ indicates that the chain gets disconnected (i.e., $J_i=0$ for the cell generated)}
\label{tab1}
\end{table*}
\end{center}

A few example $p_i/2$-decimation rules are listed in Table (\ref{tab2}). Decimating a vertical bond involves replacing the cell to its left and the cell to its right by a new cell carrying a new coupling, along with renormalization of the left- and right-neighboring vertical bonds. Again, the adjoining figure illustrates this. Some of the rules will involve $p_{i\pm 1}$ gaining an imaginary component of the order of $J_{i\pm 1}^2/p_i$, but we will restrict ourselves to scenarios where this complication plays no role.

\begin{center}
\begin{table*}
 \includegraphics[width=0.7\linewidth]{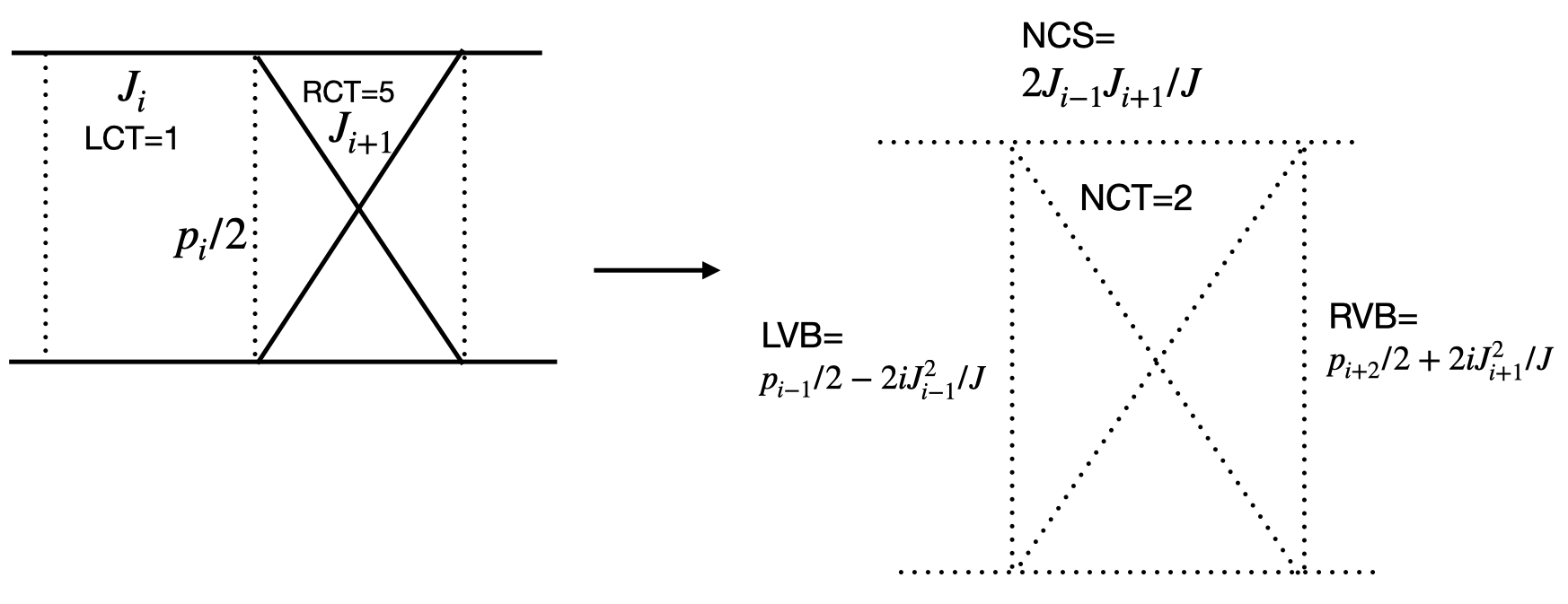}
\begin{tabular}{ |c|c|c|c|c|c|} 
 \hline
 \textbf{LCT} & \textbf{RCT} & \textbf{NCT} & \textbf{NCS} & \textbf{LVB} & \textbf{RVB} \\ 
 \hline
 1 & 1 & 2 & $2J_{i-1}J_{i+1}/p_i$ & $p_{i-1}/2+2J_{i-1}^2/p_i$& $p_{i+1}/2+2J_{i+1}^2/p_i$\\ 
 2 & 5 & -1  & 0 & $p_{i-1}/2$ & $p_{i+1}/2+8J_{i+1}^2/p_i$\\ 
 2 & 2 & 2 & $8J_{i-1}J_{i+1}/p_i$ & $p_{i-1}/2+8J_{i-1}^2/p_i$ & $p_{i+1}/2+8J_{i+1}^2/p_i$\\
 \hline

\end{tabular}
\caption{A few rules for Decimating $p_i/2$; Abbreviations: LCT=Left Cell Type, RCT= Right Cell Type, NCT= New Cell Type, NCS= New Coupling Strength, LVB=Left Vertical Bond, RVB=Right Vertical Bond. A cell type of $-1$ is a stand-in for the chain getting disconnected (via a $J_i=0$). Decimating the strongest bond in this case causes its two surrounding cells to coalesce in a new cell with updated bond strengths.}
 \label{tab2}
\end{table*}
\end{center}

We have a total of $5^3+5^2=150$ different decimation rules for RSRG. The rest of this work studies some properties of this system.

\section{RSRG Fixed Points}
\label{fp}
\begin{figure*}  
\includegraphics[width=\textwidth]{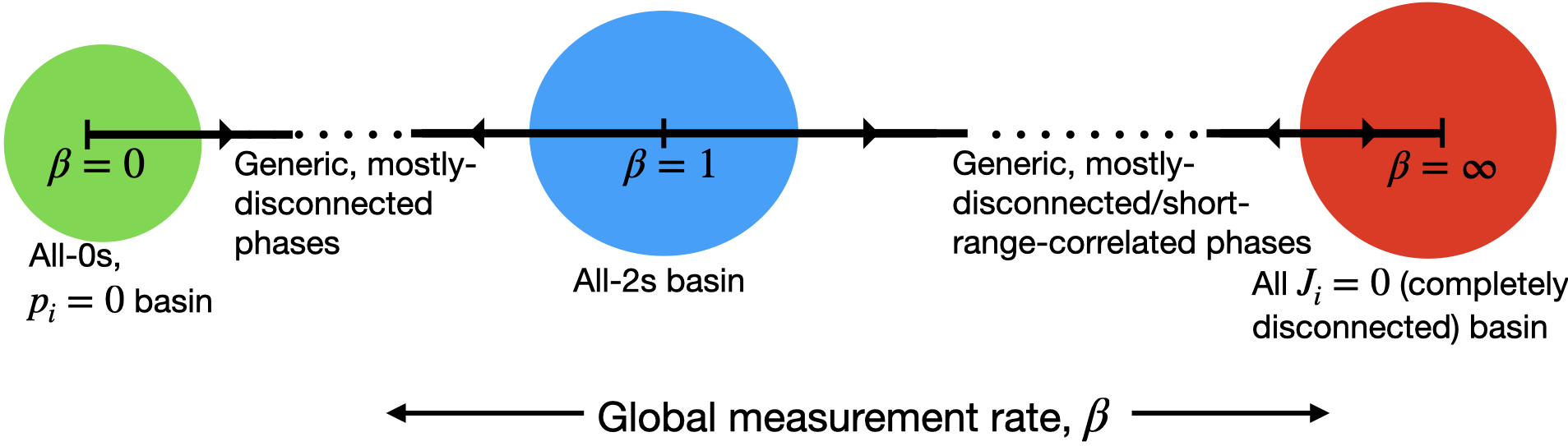}
    \caption{Phase diagram illustrating how tuning a global measurement rate $\beta$ (a bookkeeping parameter defined precisely in Fig.(\ref{fig6})(b)) can interpolate between the all-0s, all-2s, and all-disconnected phases. The all-0s IDFP and Griffiths phases (not shown) lies within the all-0s `basin' illustrated in green, while the all-2s strong disorder phase (not shown) is a subset of the all-2s basin shown in blue. The green basin is the small region where the crossover length scale to mostly-disconnected physics is larger than the system size.} The completely disconnected Zeno basin (trivial) is in red. Most initial distributions of couplings barring basins like the ones shown result in the chain ending up heavily disconnected upon RSRG, leading to only short range correlations. The arrows schematically indicate the direction of the flow of the distribution of horizontal couplings under RG flow. The immediate neighborhoods of $\beta=0$, $\beta=1$, and $\beta=\infty$ are expected to describe the same physics as $\beta=0$, $\beta=1$, and $\beta=\infty$ respectively, while larger deviations ought to flow away from the basins (at least as far as the distribution of $J_i$s goes).
    \label{fig5_5}
\end{figure*}

The RSRG scheme detailed thus far has both strong-disorder Griffiths and infinite disorder fixed points (IDFP). Infinite disorder fixed points are characterized by broad distributions of couplings that render perturbation theory asymptotically exact in their vicinity. The infinite disorder fixed points we shall analytically derive will turn out to follow from converting the RSRG rules into integro-differential equations and then finding `fixed point' solutions to those equations in the exact same way the analysis is carried out for the usual XX and Ising models \cite{rsrgMain,rsrgOrigin,rsrgAdd2}. We describe several fixed points in this section and refer the reader to \cite{rsrgMain} for the solution of the requisite integro-differential equations. The all-0s IDFP and Griffiths phases we find in the non-measured case follow from calculations identical to the ones needed for the non-measured disordered XX chain. However, a difference from the non-dissipative scenario is that these phases and fixed points describe the fastest oscillating modes of the density matrix rather than the lowest energy ones. While the all-0s infinite-disorder fixed point and Griffiths phases mathematically map to phases of the ground state in the non-measured spin chain (as in they share the same distributions of couplings and such), and the completely disconnected (all $J_i=0$) Zeno phase discussed is trivial, the strong-disorder all-2s phase is new to our knowledge, and does not appear to map to a known phase of the ground state in the non-measured scenario. The strong-disorder all-2s phase is a subset of the larger all-2s basin: the decimation rules ensure that once the chain is composed of only type-2 cells, decimations cannot introduce any other cell type. Meanwhile the all-0s IDFP and Griffiths phases are a subset of the basin consisting of only type-0 cells with all the dephasing rates set to 0. Generic initial conditions, meanwhile, lead to mostly disconnected chains with short-ranged correlations and order. Fig.(\ref{fig5_5}) illustrates these basins on a single phase diagram, given couplings initially tuned in a manner that will be discussed shortly (see Fig.(\ref{fig6})(b)). In the context of the phase diagram, the bookkeeping parameter $\beta$ (illustrated in Fig.(\ref{fig6})(b)) is defined to multiply every single dephasing rate in the initial spin chain, so tuning it can help scale  the measurement rate at every site in the system. Under RG flow, the distribution of horizontal couplings flows as indicated by the arrows and discussed in the following subsections. The phase diagram illustrates how dialing up the measurement strengths (which are tuned wrt each other in a particular fashion) causes measurement-introduced phase transitions between the basins.

\subsection{The All 0s, $p_i=0$ basin}
With all type-0 cells and $p_i=0$, vertical bonds cannot be generated via decimation. The upper and lower XX chains remain uncoupled, with couplings of equal magnitude and opposite sign, and get decimated in tandem. The maximum-oscillation-frequency modes in this case are obtained by putting the upper and lower chains in their respective ground states, i.e. the upper chain in the ground state of the original, unmeasured system and the lower chain it the (complementary) most-excited state. Phases of the fastest-oscillating coherence of the dissipationless disordered spin chain are therefore precisely those of its ground state.

In a picture such as that described by Fig.(\ref{fig6})(b), $p_i=0$ corresponds to multiplying all the vertical couplings, however they may be tuned, with the common factor $\beta=0$. If $\beta$ is ever so slightly perturbed around this point (in a region $0<\beta<\delta$ for some small $\delta$), there is a region within which the vertical couplings remain small enough that they're still never decimated; the decimation of $J_i$s proceeds exactly as in the $p_i=0$ case. In this region of the phase space, the distribution of $J_i$s thus flows exactly as it would with $\beta=0$. Outside this region, some of the vertical bonds get large enough that they too get decimated under RSRG. In the presence of such a (still small) $\beta>\delta$, the evolution of the distribution of $J_i$s splits off from evolution of the distribution in the $\beta=0$ scenario: Initially, it might follow the same trajectory, but as decimation proceeds and $J_i$s get smaller, the small vertical couplings due to the small-but-nonzero $\beta$ get more likely to be decimated and cause the evolution of horizontal couplings  to diverge from the corresponding trajectory for $\beta=0$. Therefore, outside a tiny region, we expect the all-0s basin to be repulsive in $\beta$ for the distribution of horizontal couplings (i.e. small deviations from $\beta=0$ that are larger than $\delta$ are relevant). While the size of this region, $\delta$, would be $0$ in the thermodynamic limit since the $J_i$s would eventually become smaller than any non-zero measurement rate, for finite-sized systems, the cross-over length scale at which $p_i$s begin to dominate would be larger than the system size for $0<\beta<\delta$. In other words, the all-0s fixed point is repulsive in $\beta$, but a small cross-over region around the fixed point describes the same physics as the fixed point for finite system sizes.

We briefly review the infinite disorder fixed point and the Griffiths phase hosted by this basin. The parameter $\beta$ is not made explicit in calculations; we separated it out from the vertical couplings only for aid plotting a phase diagram. 

\subsubsection{Infinite Disorder Fixed Point}
We have an IDFP when all the cells are type 0, with the vertical bonds $p_i=0$: the XX model with no measurements/dissipation. The decimation rule for $J_i$ with type-0 cells replaces $(J_{i-1},J_i,J_{i+1})\to J_{i-1}J_{i+1}/J_i$, without generating any vertical bonds. Meanwhile, since the vertical bonds are $0$, they're never decimated anyway.

We shall work in terms of $\Gamma=-\ln\Omega$, where $\Omega$ is the current largest coupling at a certain step in the RG procedure, and re-express the couplings in terms of $\zeta_i\equiv\ln(\Omega/J_i)$. In these terms, the decimation rule reads $(\zeta_{i-1},\zeta_i,\zeta_{i+1})\to \zeta_{i-1}+\zeta_{i+1}-\zeta_i$.

Let's work out how the distribution of couplings $p_J(\zeta;\Gamma)$ changes as the largest coupling is decimated (i.e., $\Omega\to\Omega-d\Omega$ or $\Gamma\to\Gamma+d\Gamma$, where $d\Gamma=-d\Omega/\Omega$). The change in $p_J(\zeta;\Gamma)$ under decimation has two sources: The redefinition of $\Gamma$ impacting the values of $\zeta_i$s, and the introduction of a new coupling correlated with the decimated ones.

We express this as \begin{multline}
    p_J(\zeta;\Gamma+d\Gamma)-p_J(\zeta;\Gamma)=p_J(\zeta+d\Gamma;\Gamma)-p_J(\zeta;\Gamma)\\+d\Gamma\,p_J(0;\Gamma)\times\\\int_0^\infty d\zeta_L\int_0^\infty d\zeta_R \delta(\zeta-\zeta_L-\zeta_R) p_J(\zeta_L;\Gamma) p_J(\zeta_R;\Gamma).
\end{multline}

The first two terms on the right hand side capture the effect of redefining $\zeta_i$s, while the integral over the left and right neighboring couplings ($\zeta_L$ and $\zeta_R$) captures the effect of decimation introducing a new coupling through the decimation rule encapsulated within the Dirac delta. We also need to renormalize the distribution due to the decimation of a coupling and the addition of a new coupling, but in the $p_i=0$ case, these two effects cancel out exactly.

This yields the integro-differential equation, 
\begin{multline}
        \frac{\partial p_J}{\partial\Gamma}=\frac{\partial p_J}{\partial\zeta}+\\p_J(0;\Gamma)\int_0^\infty d\zeta_L\int_0^\infty d\zeta_R \delta(\zeta-\zeta_L-\zeta_R) p_J(\zeta_L;\Gamma) p_J(\zeta_R;\Gamma).\label{oldidfp}
\end{multline}

If we express $p_J(\zeta;\Gamma)\equiv Q(\eta,\Gamma)/\Gamma$ using $\eta\equiv\zeta/\Gamma$ and re-express this equation in terms of $Q(\eta;\Gamma)$, we find our desired fixed point as the solution to $dQ/d\eta=0$. (So the IDFP is really only an almost-fixed point for the distribution of couplings themselves: the distribution $p$ gets scaled by $\Gamma$ as $\Gamma$ increases, but $Q$ remains fixed). As shown in \cite{rsrgMain}, the IDFP we finally obtain solving this equation is $p_J(\zeta;\Gamma)=p_\text{IDFP}(\zeta;\Gamma)\equiv\frac{\exp(-\zeta/\Gamma)\Theta(\zeta)}{\Gamma}$. In terms of the couplings $J_i$ and the largest $J_i$ at some stage, $\Omega$, this corresponds to the distribution $P_J({J};\Omega) = \frac{-1}{\Omega\ln\Omega} \left( \frac{\Omega}{{J}} \right)^{1+1/\ln\Omega} \theta(\Omega - {J})$. As the decimation progresses, $\Omega$ falls and $\Gamma$ increases, and the distribution of couplings gets broader without limit.

Interpreting more physically, with $p_i=0$, the ladder splits into two XX chains. Our search for the fastest oscillations means we seek the ground state for the ladder: choosing the most negative eigenvalues for both the upper chain (corresponding to the kets) and the lower one (the bras). The lower chain is identical to the upper chain except that the couplings all have their signs flipped. So the ground state for the lower chain is really the same as the most excited state of the upper chain. Meanwhile the ground and the most excited states are related by the antisymmetry of the spectrum: the map $\prod_{i=1}^{N/2}Z_{2i}$ that flips half the spins and negates the energy. 

The mode left behind by the procedure is described by spin singlets of arbitrarily long lengths on the upper section of the ladder, and spin triplets of arbitrarily long length in the lower section. The physics of these sections is captured by the random singlet phase observed for the ground state of a non-measured ladder: Correlations at this critical point exhibit power-law decay, with average correlators between spins initially at positions $i$ and $j$ dropping off as $1/|i-j|^2$~\cite{rsrgMain}. There is no long-range order. Interpreting the correlators in the original Hilbert space requires a bit more care since they describe correlations in the oscillations of coherences in the density matrices rather than static correlations in an eigenstate. The entanglement entropy of a section of length $L$ scales as $S_L\sim\ln L$, as expected for a critical phase~\cite{entang}.

\subsubsection{Griffiths (Random Dimer) Phase}
The ground state of the non-measured XX spin chain~\cite{rsrgReviewOld,Griffiths2,Griffiths3}  has a Griffiths phase described by the off-critical region around infinite-disorder criticality. It is described by dimerization of the chain: the even and odd bonds lie on different distributions, and the distance from criticality (i.e. the random singlet phase) is characterized by the difference between the distributions. The dynamical exponent is non-universal and varies continuously over this phase. The Griffiths phase has long-range order. 

Since our RSRG rules for type-0 cells are precisely the same as the ground-state-RSRG rules for the non-measured disordered XX model (i.e., $(J_{i-1},J_i,J_{i+1})\to J_{i-1}J_{i+1}/J_i$), the random dimer phase must precisely carry over to our description of fast-oscillating-coherences. More explicitly, since $p_i=0$ means we are studying the fastest oscillating mode of a spin ladder constituting of two decoupled chains, we could envision having the upper chain of the ladder in the ground state of this random dimer phase, and the lower chain in its most excited state (as mapped by $\prod_{i=1}^{N/2}z_{2i-1}$, flipping the spectrum) and arrive at an off-critical random dimer phase for the fastest oscillating mode in the density matrix.
\subsection{The All 2s Basin}
If at any point all the cells in the ladder are of type $2$, the decimation rules ensure that they continue to remain of type $2$. 

\begin{figure}  
\begin{subfigure}{\linewidth}
\caption{}
    \includegraphics[width=\linewidth]{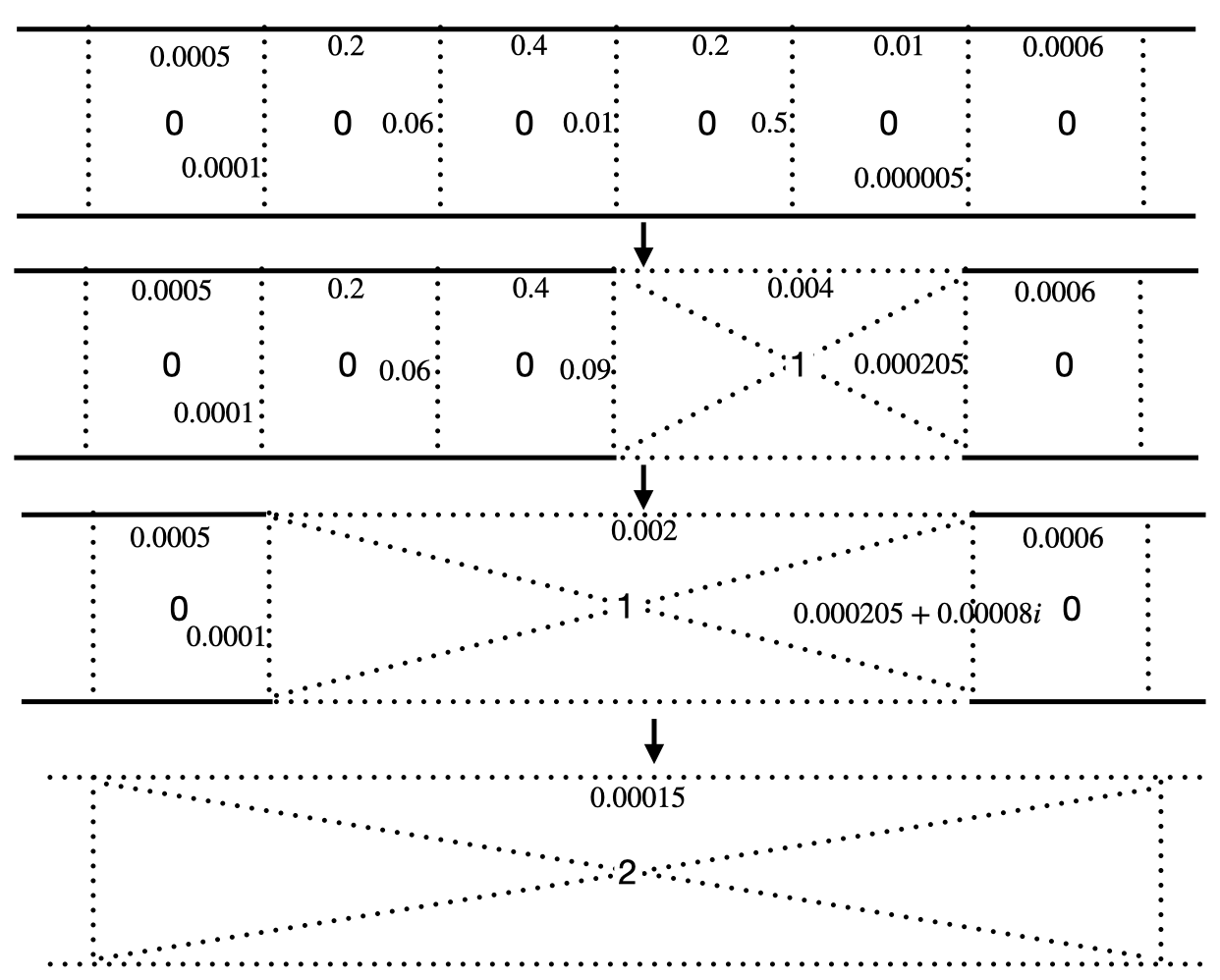}
    
  \end{subfigure}
  \begin{subfigure}{\linewidth}
  \caption{}
    \includegraphics[width=\linewidth]{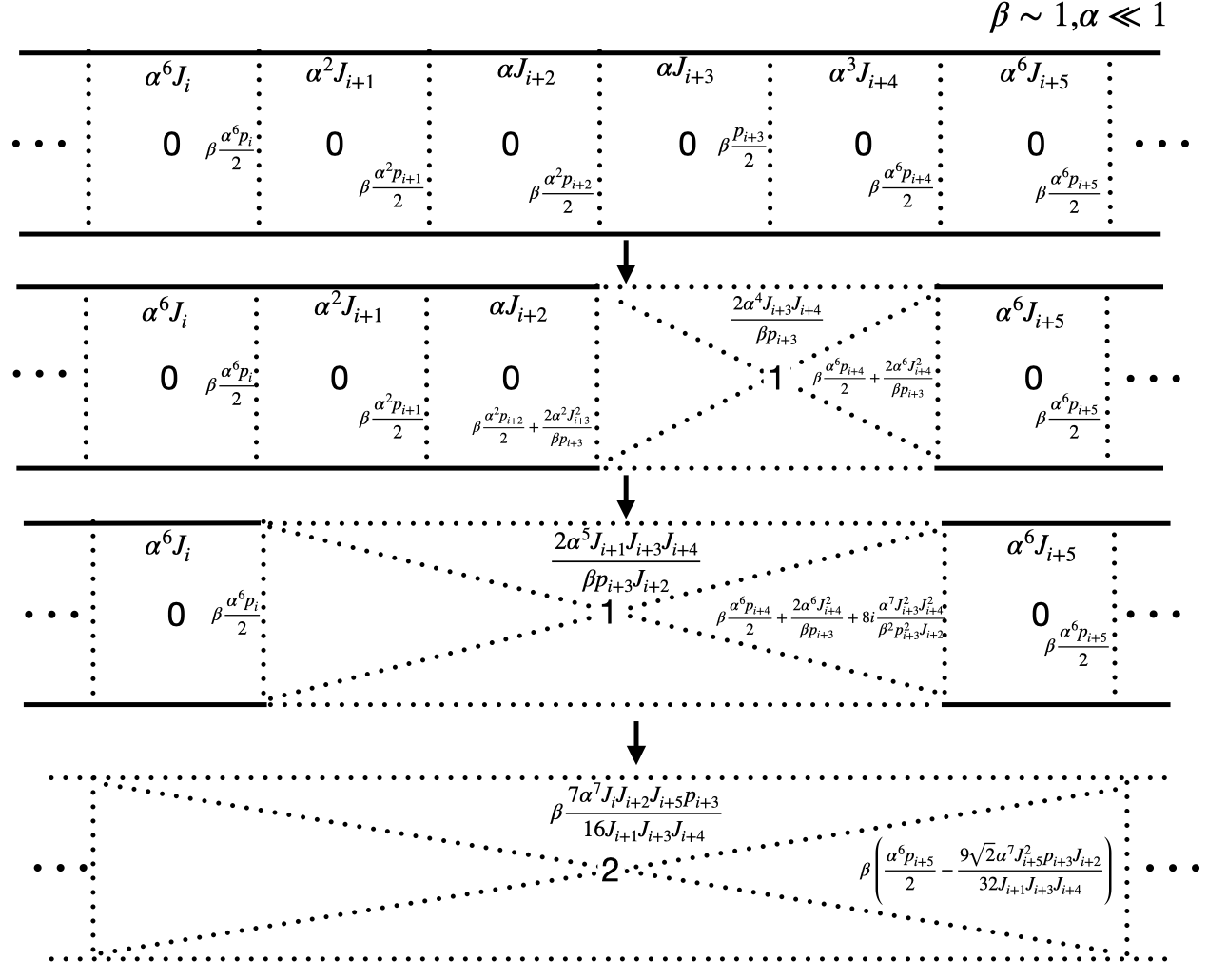}

  \end{subfigure}
    \caption{(a) A sequence of six initial type-0 cells (no diagonal couplings) can coalesce via RSRG decimations into a single type-2 cell. This transition indicates that even simple initial conditions can evolve into the all-2s basin, and ultimately the all-2s strongly-disordered fixed point. (b) The same process, illustrated using the bookkeeping parameters $\alpha\ll 1$, used to control the relative sizes of couplings, and $\beta$, which globally tunes the measurement/dephasing rate and is set to be $\sim 1$ here.}
    \label{fig6}
\end{figure}

We describe a possible route that a chain that is initially all-0s (no diagonal couplings) might take to reach an all-2s basin. Observe for instance Fig.(\ref{fig6})(a). It shows how a clump of 6 type-0 cells can form a type-2 cell. In general, therefore, there exist initial conditions that take us to the all-2s basin starting from an initial all-0s state, justifying our study of the all-2s IDFP. Notice how some of the $p_i$s appear to become complex during decimation. However, the imaginary part generated is infinitesimally small for appropriate initial conditions; moreover, such complex bonds are never decimated in the pathway to type-2 cells shown.

Fig.(\ref{fig6})(b) uses general bookkeeping parameters to illustrate the same route. The relative magnitudes of couplings are controlled by the $\alpha$, which we choose to be sufficiently small for the decimation rules to go through in the manner illustrated. We introduce $\beta$ to globally control the measurement rates (and help us plot a phase diagram, Fig.(\ref{fig5_5})), but set it equal to unity in this figure.

In the context of the phase diagram, there is a finite region in the vicinity of $\beta=1$ (say $|\beta-1|<\delta$) that does not affect the decimation scheme employed to produce type-2 cells, and therefore still results in the production of such cells. Once the entire chain has been reduced to type-2 cells formed by the annihilation of sextuples of type-0 cells, $\beta$ serves to uniformly scale every coupling, horizontal and vertical, in the resulting chain; the horizontal/diagonal couplings for the type-2 cell generated are $J_i'=\beta \frac{7 \alpha^7 J_{i} J_{i+2} J_{i+5} p_{i+3}}{16 J_{i+1} J_{i+3} J_{i+4}}$ while the vertical coupling is $p_i'=\beta \left( \frac{\alpha^6 p_{i+5}}{2} - \frac{9\sqrt{2} \alpha^7 J_{i+5}^2 p_{i+3} J_{i+2}}{32 J_{i+1} J_{i+3} J_{i+4}} \right)$. Any $\beta$ in the region $|\beta-1|<\delta$ leads to an all-2s chain with the same couplings, but scaled by $\beta$. The existence of a finite stability region $\delta$ is guaranteed by the hierarchy of scales introduced by the parameter $\alpha$. Since the Real-Space Renormalization Group (RSRG) flow is determined entirely by the relative magnitudes of the couplings, a uniform rescaling by $\beta$ leaves the trajectory of the flow invariant. The distribution of $\zeta$s thus remains the same for any $|\beta-1|<\delta$. Once in this basin, the decimation rules ensure the system cannot exit, as type-2 cells generate only other type-2 cells under decimation.

Outside the range of $\beta$ that leads to type-2 cells, long range order generically breaks down, we no longer land up in the all 2s phase, and the chain ends up mostly disconnected, with only small connected parts. Hence, we expect sizable deformations from the $\beta=1$ point to be repulsive. 

Note again that we do not use the parameters $\alpha$ and $\beta$ in calculations; they were only made explicit in Fig.(\ref{fig5_5})) and Fig.(\ref{fig6}) for illustrative purposes. We discuss a family of fixed points with arbitrarily strong disorder that occur in this basin.

\subsubsection*{Strong Disorder Phase}
Observe the decimation rules for this case: $(J_{i-1},J_i,J_{i+1})\to 7J_{i-1}J_{i+1}/(2J_i)$ and $(J_{i-1},p_i,J_{i+1})\to 8 J_{i-1}J_{i+1}/p_i$, with renormalization of vertical couplings via addition of terms of the order of $J_{i\pm 1}^2/p_i$ and $p_{i\pm 1}^2/J_i$. The decimation under consideration can be made to look the same if we rescale $p_i\to 7/16 p_i$. Thus $c_{2i}=J_i$, $c_{2i-1}=7p_i/16$. Then the RSRG rule for $J_i$-decimation is $(c_{2i-2},c_{2i-1},c_{2i},c_{2i+1},c_{2i+2})\to 7c_{2i-2}c_{2i+2}/c_{2i}$, while the RSRG rule for $p_i$ decimation is  $(c_{2i-1},c_{2i},c_{2i+1})\to 7c_{2i-1}c_{2i+1}/(2c_i)$.

 If the largest $c_i$ at some stage is $\Omega$, define again $\Gamma\equiv -\ln\Omega$, and $\zeta_i=\ln(\Omega/c_i)$. Call the distribution of $J_i$s, $p_J(\zeta;\Gamma)$, and the distribution of $p_i$s, $p_p(\zeta;\Gamma)$. The equation for the evolution of the $p_i$-distribution reads,
 \begin{multline}
        \frac{\partial p_p(\zeta;\Gamma)}{\partial \Gamma} = \frac{\partial p_p(\zeta;\Gamma)}{\partial \zeta} + p_p(0;\Gamma) p_p(\zeta;\Gamma)\\+2(p_p(0,\Gamma)+p_J(0,\Gamma))\times\\\int_0^\infty d\zeta_L\int_0^\infty d\zeta_R \delta(\zeta-h(\zeta_L,\zeta_R))p_p(\zeta_L)p_J(\zeta_R)\\-2p_p(\zeta,\Gamma)(p_p(0,\Gamma)+2p_J(0,\Gamma)),\label{pidfp}
\end{multline}
with $h(\zeta_L,\zeta_R)\equiv\zeta_L-\ln\left[1+\exp(\zeta_L-2\zeta_R)\right]$.
Here the second term serves to re-normalize the distribution after the largest $p_i$ is decimated. 

The third term specifies the renormalization rule for the neighboring vertical bonds when horizontal or vertical bonds are decimated, via the function $h(\zeta_L,\zeta_R)$. The fourth term accounts for the renormalization of the distribution due once these new neighboring vertical couplings have been generated. Note that we have neglected constant multiplicative factors in the renormalization rule for the vertical coupling since they only contribute logarithms of small numbers to $h(\zeta_L,\zeta_R)$, and we are eventually interested in the regime of broad distribution, i.e., typically large $\zeta$s.

 Meanwhile, the flow of the $J_i$ distribution is described by,
 \begin{multline}
\frac{\partial p_J(\zeta;\Gamma)}{\partial \Gamma} = \frac{\partial p_J(\zeta;\Gamma)}{\partial \zeta} - p_p(0;\Gamma)p_J(\zeta;\Gamma)\\+ \left(p_J(0;\Gamma) + p_p(0;\Gamma)\right) \times\\\int_0^\infty d\zeta_L \int_0^\infty d\zeta_R \, \delta(\zeta - \zeta_L - \zeta_R-\ln(7/2)) \, p_J(\zeta_L;\Gamma) p_J(\zeta_R;\Gamma),\label{Jidfp}
\end{multline}
where the third term has been modified as compared to Eq.\eqref{oldidfp} to include the decimation of $p_i$s. The second term arises from the need to re-normalize the distribution after the addition of this effect. Again, close to an infinite disorder fixed point, $\zeta$s are typically large, and hence $\ln(7/2)$ can be ignored in the Dirac delta.

\begin{figure}
\begin{subfigure}{0.9\linewidth}
    \caption{}
    \includegraphics[width=\linewidth]{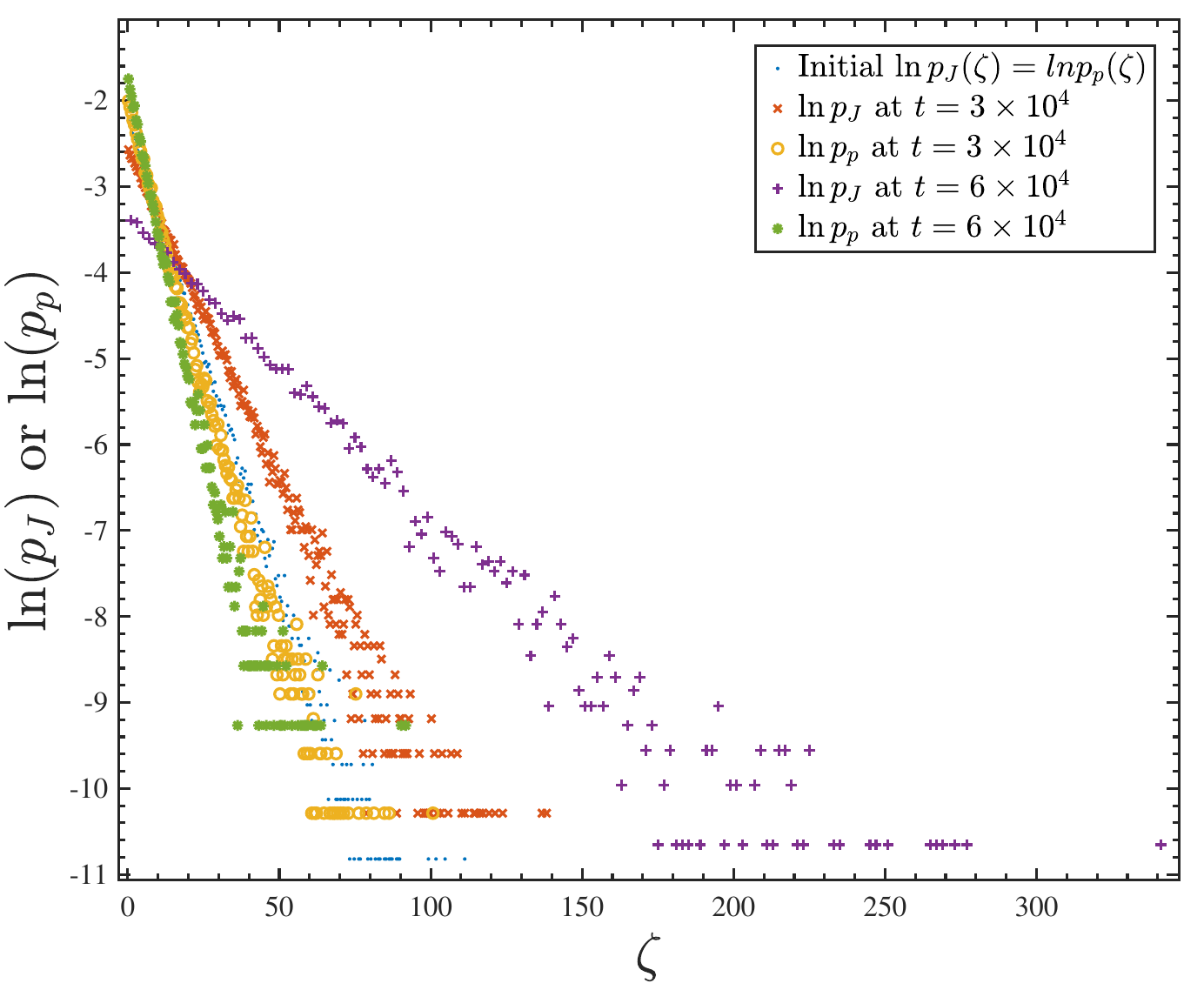}
  \end{subfigure}
  \begin{subfigure}{\linewidth}
      \caption{}
    \includegraphics[width=\linewidth]{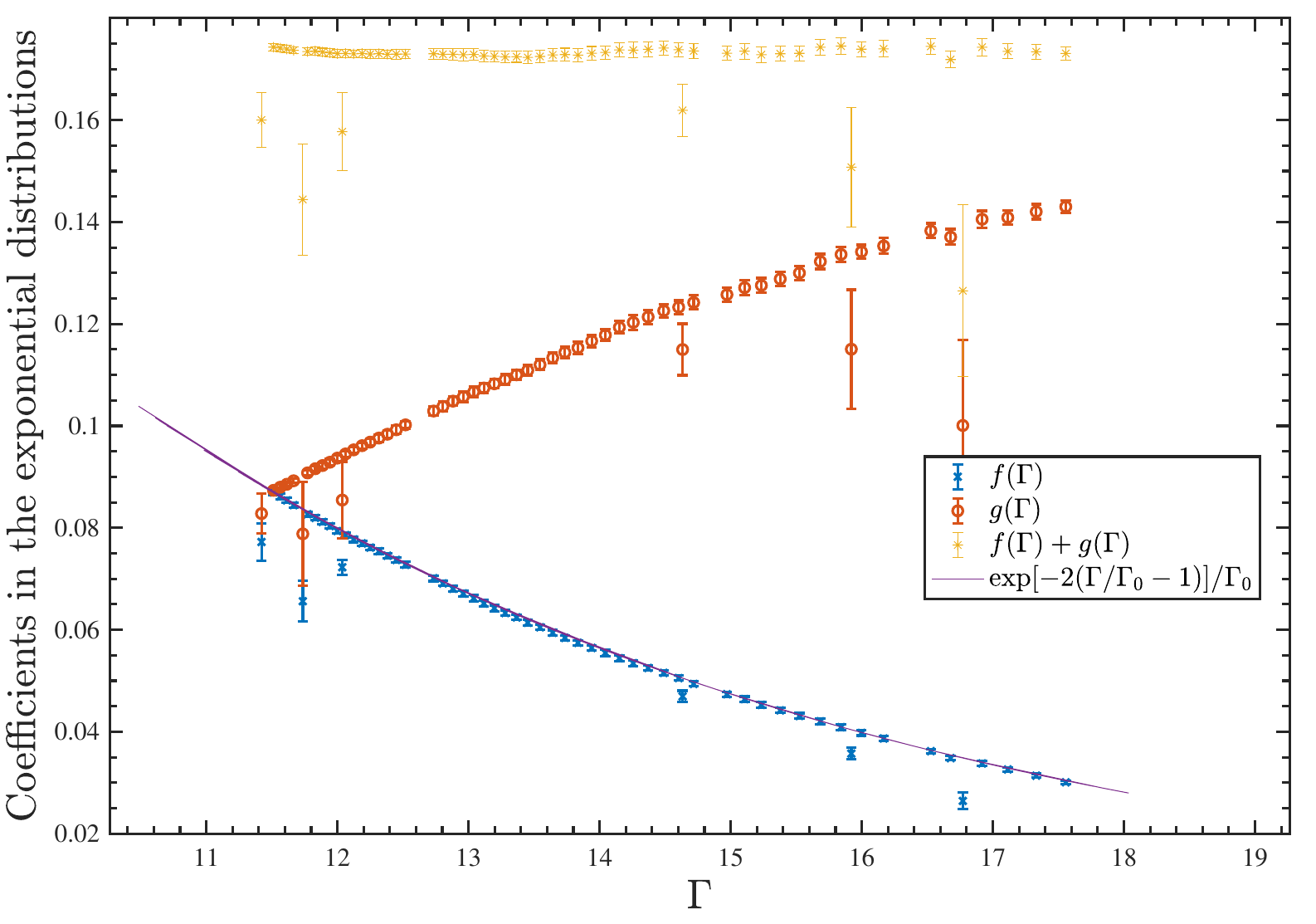}
  \end{subfigure}
    \caption{Simulating RSRG on an $N=10^5$ all-2s ladder: (a) Broadening of coupling distributions observed for initial conditions $p_p(\zeta)=p_J(\zeta)=\exp(-\zeta/\Gamma_0)$, by plotting $\ln p_p(\zeta)$ and $\ln p_J(\zeta)$ vs $\zeta$ at $t_\text{RSRG}=0$ ($\Gamma_0=9.2$), $t_\text{RSRG}=3\times 10^4$ ($\Gamma=10.8$), and $t_\text{RSRG}=6\times 10^4$ ($\Gamma=14.3$) RSRG time steps-- Straight lines justify the exponential ansatze used to solve the RG flow equations, (b) With $\Gamma_0=11.5$, $f(\Gamma)$ and $g(\Gamma)$ as extracted from fitting exponential distributions for the couplings, along with error bars. Note that $f(\Gamma)+g(\Gamma)$ is roughly constant numerically. Furthermore, the analytical prediction for $f(\Gamma)$ collapses onto the numerical plot.}
    \label{Fig4}
\end{figure}

To (approximately) solve the equations, we use 
the ansatze
\begin{align}
    p_p(\zeta;\Gamma) &= g(\Gamma)\exp(-g(\Gamma)\zeta)\\
    p_J(\zeta;\Gamma) &= f(\Gamma)\exp(-f(\Gamma)\zeta).\label{ansatz}
\end{align}
These could be justified numerically: Fig.(\ref{Fig4})(a) shows how the distributions for $p_i$s and $J_i$s separate out but remain roughly exponential.
 Plugging into Eq.\eqref{Jidfp} and comparing the coefficients of the $\exp(-g(\Gamma)\zeta)$-term and the $\zeta\exp(-g(\Gamma)\zeta)$-term consistently yields,
 \begin{equation}
     \frac{df}{d\Gamma}=-fg-f^2,\label{df}
 \end{equation}
 where we've dropped the $\Gamma$-dependence for simplicity.
 But plugging the ansatze into  Eq.\eqref{pidfp} gives us for positive $\zeta$,
  \begin{equation}
     \frac{dg}{d\Gamma}-\zeta g\frac{dg}{d\Gamma}=-2g(f+g)+fg(f+g)\exp(-f\zeta/2)B(g,f/2),
 \end{equation}
 in terms of the Beta function $B(g,f/2)$. Simplifying using $B(x,y)=\Gamma(x)\Gamma(y)/\Gamma(x+y)$ and $\Gamma(x)\approx 1/x$ for $x\ll 1$, in the limit $f\ll g\ll 1, f\zeta\ll 1$,
   \begin{equation}
     \frac{dg}{d\Gamma}-\zeta g\frac{dg}{d\Gamma}\approx f(f+g)-\zeta fg(f+g)\implies \frac{dg}{d\Gamma}\approx fg+f^2.\label{dg}
 \end{equation}

 Comparing eqns.\eqref{dg} and \eqref{df}, observe that $f+g$ is approximately constant.

 Let's start with $f(\Gamma_0)=1/\Gamma_0=g(\Gamma_0)$. This violates the condition $f\ll g$ under which the approximations were derived, but if it eventually flows to $f\ll g$ (as we shall verify numerically), the approximation eventually holds. Then,
 \begin{align}
     f(\Gamma)= &\frac{1}{\Gamma_0}\exp[-2(\Gamma-\Gamma_0)/\Gamma_0],\\ g(\Gamma)=&\frac{1}{\Gamma_0}\left\{2-\exp[-2(\Gamma-\Gamma_0)/\Gamma_0]\right\}.
 \end{align}
 We verify the validity of this solution in Fig.(\ref{Fig4})(b), which shows how $f(\Gamma)+g(\Gamma)$ is approximately constant and $f(\Gamma)$ extracted from simulations collapses onto our analytical explanation.

In terms of $J_i$ and $p_i$, the probability distributions read,
\begin{align}
    P_p(p;\Gamma) & =& \frac{g(\Gamma)}{p}\left(\frac{7p}{16\Omega}\right)^{g(\Gamma)}\Theta(\Omega-\frac{7p}{16})\\
    P_J(J;\Gamma) & = & \frac{f(\Gamma)}{J}\left(\frac{J}{\Omega}\right)^{f(\Gamma)}\Theta(\Omega-J).\label{ansatz}
\end{align}

This indicates that the $J_i$-distribution gets broader exponentially faster than for the all-0s IDFP, where the width only goes down as $1/\Gamma$. This is aided by the fact that not only the decimation of the $J$-couplings, but also the $p_i$-couplings contribute to the creation of new $J_i$ values that are likely to be smaller than the typical $J_i$, and hence broaden the distribution. 

 On the other hand, the $p_i$-distribution gets narrower due to the renormalization of the vertical bonds, as it serves to slightly increase the $p_i$s. However, its width saturates. Therefore, we can choose the distribution of $p_i$s to be as broad as we please by choosing a large enough $\Gamma_0$. Meanwhile the $J_i$ values tend to broaden out rapidly by themselves. It is worth noting that if we started with initial conditions $f(\Gamma_0)\ll g(\Gamma_0)$ satisfying our approximation scheme at the outset, the width of the $p_i$-distribution would essentially remain fixed while the $J_i$-distribution continued getting exponentially broader. This is a nontrivial fixed point that the system gravitates towards for certain initial conditions.

\begin{figure}
    \centering
\begin{subfigure}{\linewidth}
    \caption{}
    \includegraphics[width=\linewidth]{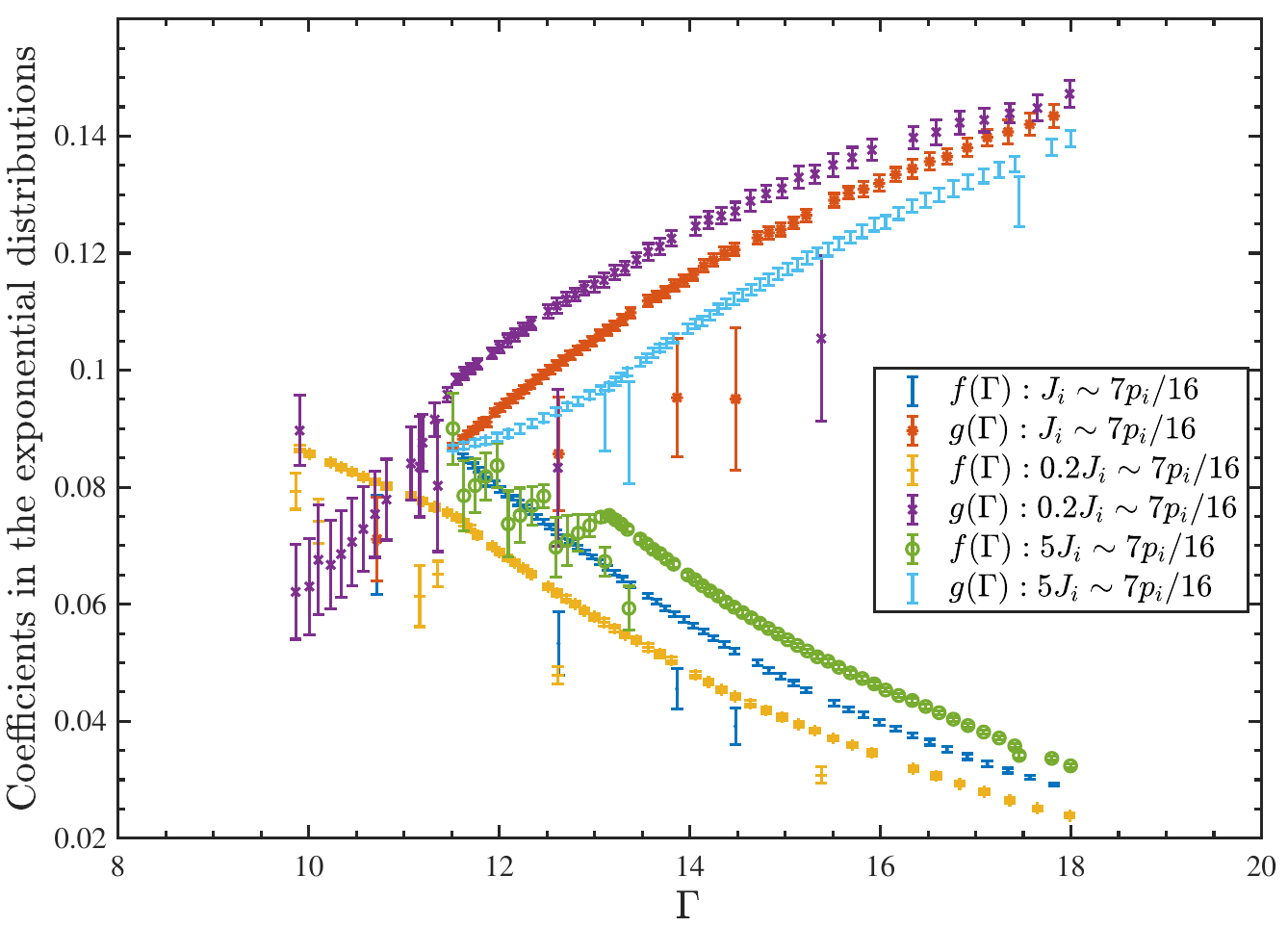}
  \end{subfigure}
  \begin{subfigure}{\linewidth}
      \caption{}
    \includegraphics[width=\linewidth]{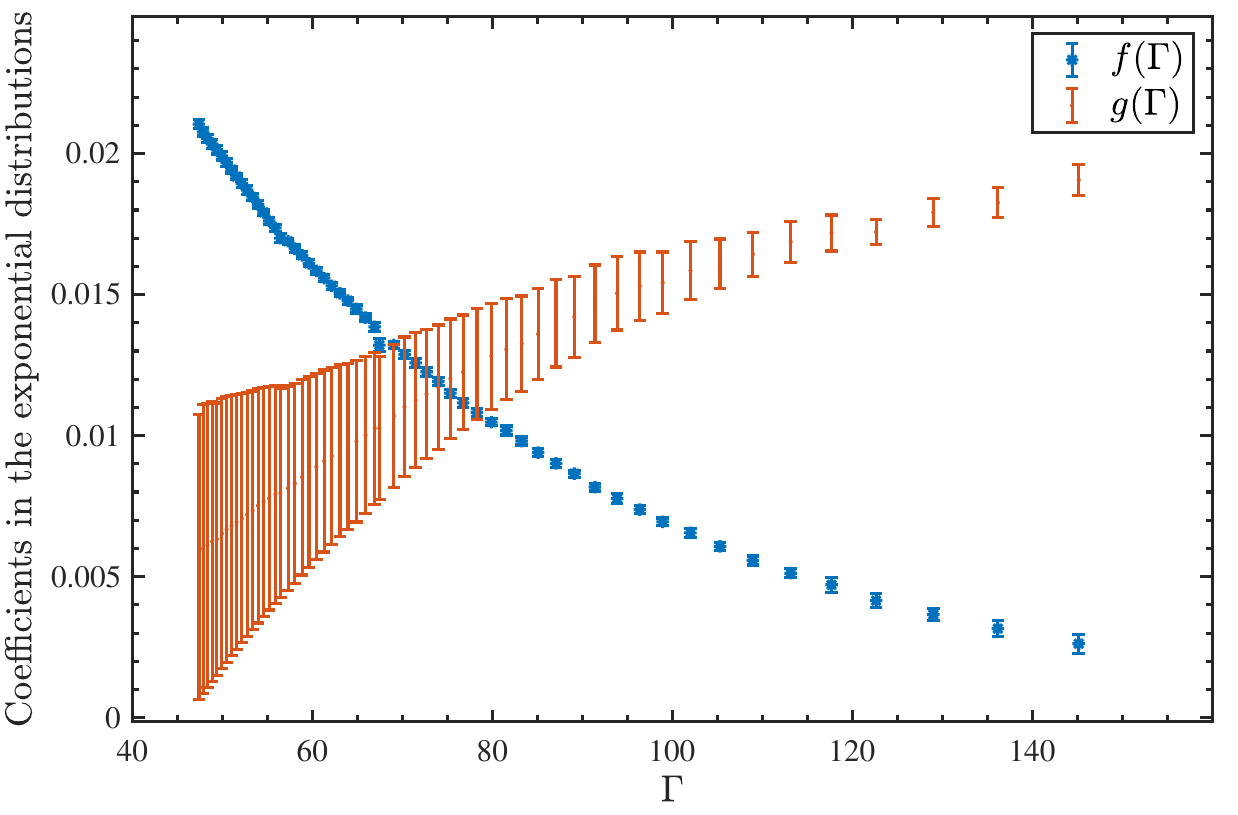}
  \end{subfigure}
    \caption{Stability and attractiveness of the fixed point for an $N=10^5$ all-2s ladder: (a) ($\Gamma_0=11.5$) Perturbing the initial distribution of $J_i$s multiplicatively and extracting $f(\Gamma)$ and $g(\Gamma)$ by fitting the distributions with exponential ansatze to observe how they still land up being strongly broadened, (b) ($\Gamma_0=46$) Starting with all $p_i=0$: observe that the distribution still tends to an exponential distribution (the error bars get smaller) with the same qualitative behavior as before: saturation of the numerically extracted $g(\Gamma)$ and decay of $f(\Gamma)$.}
    \label{Fig5}
\end{figure}

 We analytically found one set of solutions pertaining to strongly disordered fixed points: a one-dimensional manifold parametrized by the initial width $\Gamma_0$. This solution does not necessarily describe all the strongly disordered fixed points that exist, and in fact, there may be an entire basin of fixed points of which our solutions form only a lower-dimensional subspace. We now proceed to numerically simulate our decimation rules for a disordered periodic XX ladder with $N=10^5$ sites, beginning with all type-2 cells. In Fig.(\ref{Fig5}), we present numerical evidence that both small and large perturbations to the initial distribution can still lead to broadening, indicating that there is in fact an attractive basin in the vicinity of our solutions. We studied a single realization of the system starting with distributions described in the captions, fitted the distributions at intermediate time steps with exponentials (in $\zeta$) and plotted the extracted rate parameter ($f(\Gamma)$ and $g(\Gamma)$ for our ansatz above) with the error in the fit.  In Fig.(\ref{Fig5})(a), we perturb the $J_i$-distribution multiplicatively. Extracting $f(\Gamma)$ and $g(\Gamma)$, we see that we still end up getting strong broadening as $g(\Gamma)$ grows but saturates, while $f(\Gamma)$ decreases: the perturbed distributions have widths that closely follow the nonperturbed ones as shown.

\begin{figure}
    \centering
\begin{subfigure}{\linewidth}
    \caption{}
    \includegraphics[width=\linewidth]{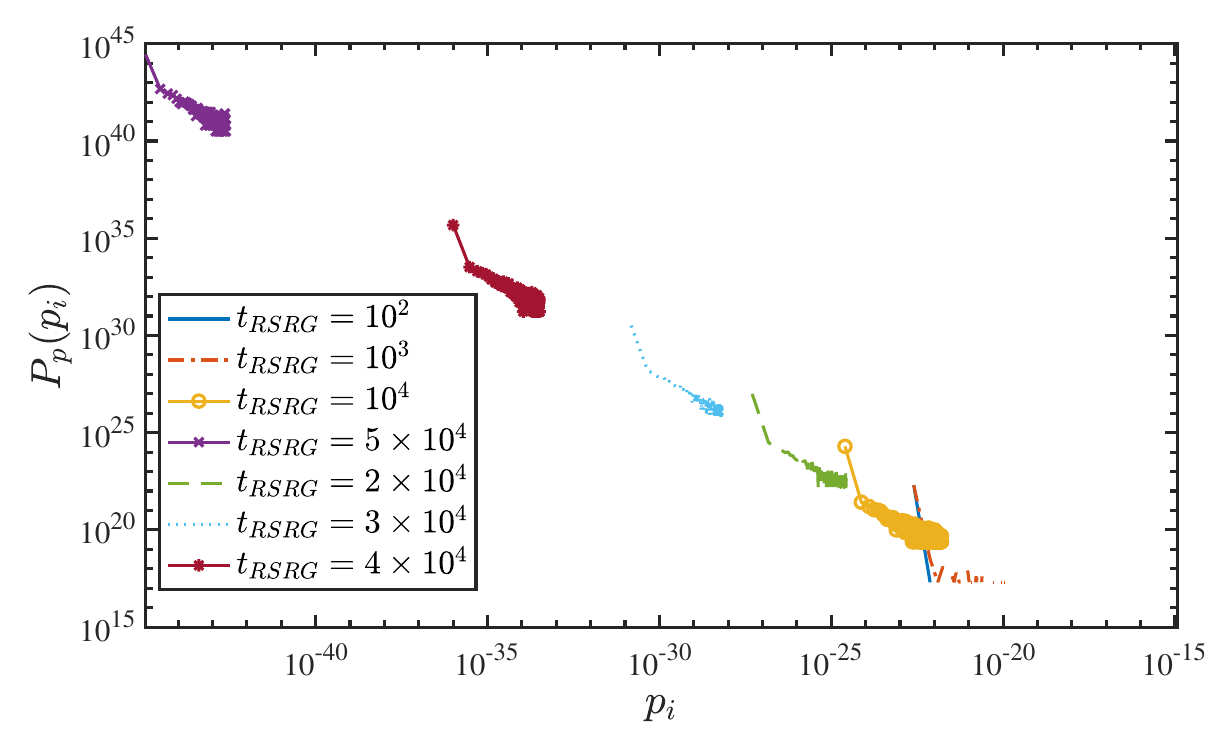}
  \end{subfigure}
  \begin{subfigure}{0.9\linewidth}
      \caption{}
    \includegraphics[width=\linewidth]{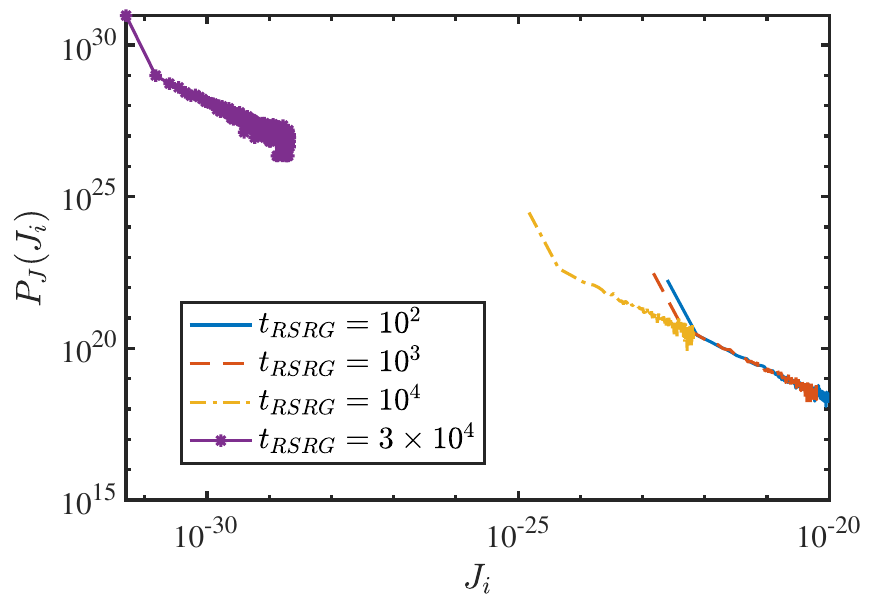}
  \end{subfigure}
    \caption{(Using $\Gamma_0=46$) Log-log plots of coupling distributions at various stages of the RSRG procedure starting from a scenario with initially zero vertical couplings (all $p_i=0$): (a) Vertical coupling distribution $P_p(p_i)$, (b) Horizontal coupling distribution $P_J(J_i)$. Both distributions visibly broaden with RSRG progression.}
    \label{Fig7}
\end{figure}

 In (b), we set all the $p_i$s initially to $0$. Again, the plot shows the numerically extracted $f(\Gamma)$ and $g(\Gamma)$ obtained by fitting an exponential distribution onto the data. The error bars for $p_i$ start out large, since the distribution is far from exponential, but get smaller as RSRG progresses. Furthermore, as with our analytical solutions, $g(\Gamma)$ appears to saturate, while $f(\Gamma)$ continues to get smaller: the distribution of $J_i$s keeps getting broader gets closer to exponential and saturates in width.  The probability distributions in terms of $J_i$ and $p_i$ at several time-steps are shown in Fig.(\ref{Fig7}): it can be seen how the distributions get broader as RSRG progresses. 

The mode generated by decimation has both pairs of entangled spins stretching over arbitrarily long length scales that oscillate and decay in a correlated manner, and lone single sites. Let's try to quantify the correlations. If there are $n(\Gamma)$ spins remaining once we've gone from $\Gamma_0$ to $\Gamma$, going up to $\Gamma+d\Gamma$ erases $(p_p(0,\Gamma)+2p_J(0,\Gamma))n(\Gamma)d\Gamma$ more spins. Hence, with $n(\Gamma_0)=N$, we find $$n(\Gamma)= N \exp\left[-\frac{2\bigl(\Gamma-\Gamma_0\bigr)}{\Gamma_0}+\frac{1}{2}(1 - e^{-2(\Gamma-\Gamma_0)/\Gamma_0})\right].$$ 

Therefore the typical length scale $L\sim N/n\sim \exp[2\Gamma/\Gamma_0]\sim\Omega^{-2/\Gamma_0}$. As decimation proceeds, large $p_i$s become more common than large $J_i$s, and $p_i$s are decimated more and more often. $J_i$s are decimated $p_J(0,\Gamma)/p_p(0,\Gamma)\sim \exp[-2(\Gamma-\Gamma_0)/\Gamma_0]\sim n/N$ as rarely as $p_i$s. Consider spins initially at positions $i$ and $j$ in the spin chain. Then if both spins survive until length scale $|i-j|$ or $n\sim N/|i-j|$, they're likely to be close/next to each other.  The probability of this happening scales as $1/|i-j|^2$. Finally, if the $J$-coupling linking the two spins is decimated, they get into an entangled state listed in the Appendix, developing significant correlations. Since the relative probability of decimating a horizontal coupling is $n/N\sim 1/|i-j| $, correlators for two spins separated by a distance $|i-j|$ ought to decay as $1/|i-j|^3$. Similar to the argument presented by Fisher~\cite{rsrgMain}, we expect the mean correlation to be dominated by these rare long range spin pairs. Correlations between typical pairs of spins are exponentially weaker. Therefore, we predict a power-law decay for correlators in this state, with an inverse cube law signifying a faster fall-off than in the non-measured (all-0s) IDFP, which had an inverse square dependence. These correlators capture not only how the spins oscillate, but also how they decay together.

A quick calculation along the lines of \cite{entang} reveals the behavior of the entanglement between parts of the ladder in this phase. The operator entanglement entropy~\cite{opee} between two parts of the ladder is measured by the number of singlets crossing over between the two parts. Let the two parts be connected by a bond $B$. Ignoring the history dependence for bond formation to get an approximate result, the number of bonds decimated as $\Gamma\to\Gamma+d\Gamma$ is $d\bar n=p_J(0;\Gamma)d\Gamma=f(\Gamma)d\Gamma$. Therefore, the average number of decimated bonds over B is $\bar n=\frac{1}{2}(1-\exp[-2(\Gamma-\Gamma_0)/\Gamma_0])$. As $\Gamma$ increases, this just saturates, indicating that the entanglement is area law. This is peculiar: the correlators falling off as $f(\Gamma)/[g(\Gamma)|i-j|^2]$ implies they fall off algebraically since even though $f$ falls exponentially with $\Gamma$, the length scale also goes up exponentially with $\Gamma$, while the entropy saturates given $f$'s said exponential decay. This could perhaps be an artifact due the entanglement being a 4-spin quantity in the non-Hermitian ladder. We also do not preclude the possibility that a more precise calculation of correlators might resolve this peculiarity.

\begin{figure}
    \centering
    \includegraphics[width=\linewidth]{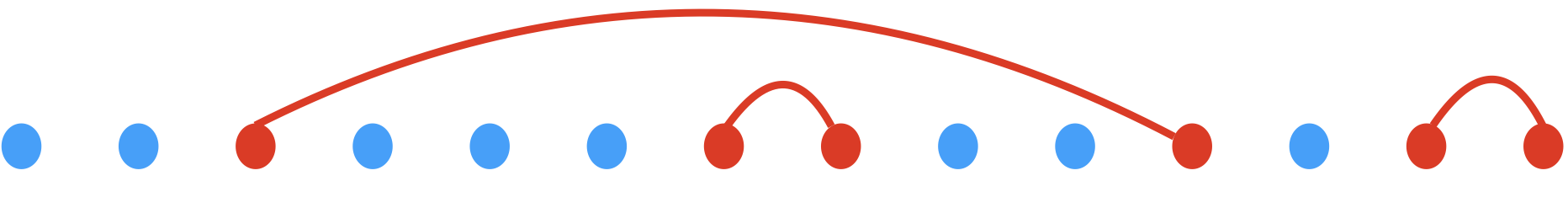}
    \caption{A schematic picture of a long-lived fast-oscillating mode that we might reach after decimations. The blue points denote sites arising due to decimated $p_i$s: these sites are in maximally mixed states. The red bonds denote entangled pairs of sites entangled due to decimation of $J_i$: decimations of $J_i$ produce entangled pairs that correlate not only spatially, but also dynamically in how they dephase together. These pairs exist at all length scales and cause long-range correlations.}
    \label{Fig7.5}
\end{figure}

Fig.(\ref{Fig7.5}) depicts the picture of how the state we reach might look. Decimating $p_i$s produced maximally mixed states, while strong $J_i$s lead to mixed states supported on pairs of sites. These pairs exist at arbitrarily long lengths, but are  fewer at long length scales than in the random singlet phase.
\subsection{The Completely Disconnected (Zeno) Basin ($J_i=0$)}
If we start with all $J_i=0$ and only $p_i\neq 0$, the $J_i$s remain $0$: the completely disconnected chain remains completely disconnected under RSRG.

A chain that for instance has its $p_i$s all larger than its $J_i$s is in this phase, as only the vertical bonds ever get decimated under RSRG. Since decimation of vertical bonds and selection of the longest lived state is tantamount to projecting local sites into maximally mixed states, the state captured by the algorithm in this case is the maximally mixed state: always a steady state of the Lindbladian. There's no entanglement at this trivial fixed point.

In Fig.(\ref{fig5_5}), $\beta=\infty$ captures this basin. Upon decimation, the couplings $J_i$ all flow to 0. Deviations from this point, i.e. large but finite $\beta$, simply cause the distribution of $J_i$s to flow back to this point. If all the vertical couplings are significantly larger than all the horizontal ones, the vertical ones keep getting decimated and the horizontal ones generated under decimation get closer and closer to 0. Hence, we expect the point $\beta=\infty$ to be attractive for the flow of the $J_i$-distribution.

Larger deviations from $\beta=\infty$ ought to be repelled, since as the vertical couplings get smaller, the horizontal couplings no longer flow to 0 when the vertical bonds are decimated; moreover, the reduction of $\beta$ implies eventually horizontal couplings begin to get decimated, and while we probably still have only short-ranged correlations, the system is no longer completely disconnected.

\section{Conclusion}
We studied long-lived fast-oscillating modes of disordered XX spin chains with random measurements and randomly picked measurement probabilities, mapping the problem to a non-Hermitian XX spin ladder. The ladder could then be attacked using RSRG, resulting in a set of decimation rules that we both simulated numerically and studied analytically. In addition to phases expected from the non-measured scenario (where maximizing oscillation frequency is tantamount to looking at states with extremal energy and yields the usual phases obtained for the non-measured ground state), we found a phase with long-range order that exists only in the presence of measurements: our `all-2s' phase. We obtained analytical solutions for a set of strong-disorder fixed points within this phase and conjectured area law entanglement and algebraically decaying correlators. Tuning a global measurement rate could take us between the non-dissipative all-0s, the all-2s, and the Zeno basins, demonstrating the effect of competition between unitary evolution and dephasing in this disordered setting.

In this paper, we restricted ourselves to studying cases where the imaginary couplings could be neglected. For instance, for the all-2s case, the path we delineated to obtain type-2 cells starting from type-0 cells in did involve the generation of complex couplings-- However, in the later steps, these couplings disappeared without ever having been decimated. In general, we might expect the fixed point structure to be altered by the imaginary components in cases where they contribute to large $J_i$s or $p_i$s that end up getting decimated. A more complete analysis for complex $J_i$ and $p_i$ would be a natural future step.

Some further applications of our method are relatively straightforward. The analysis we performed could also be carried out to yield information about between $2^{N/2}$ to $2^N$ modes in the spectrum out of the total of $2^{2N}$ total modes that the Hamiltonian hosts, by opting to pick the shortest-lived or highest-positive-frequency eigenstates in the RSRG-X routine. Obtaining a stable distribution of couplings might necessitate being systematic about how we pick our branches. Usually, the branches are sampled probabilistically, and a thermal distribution is used for sampling the eigenstates. However, since we deal with open system modes rather than closed system energy eigenstates, one would need to consider other natural distributions to sample.

The all-0s and all-2s fixed points would also show up if we tried to perform RSRG on a (non-measured, Hamiltonian) reflection-symmetric disordered XX spin ladder: two identical disordered XX spin chains coupled to each other via rungs carrying random XX couplings. Effectively, we thus also derived RSRG-X for and studied the physics of the ground and highest excited state manifolds of disordered reflection-symmetric XX spin ladders. Preliminary calculations show that all-2s sink turns out to be a lot easier to reach in this scenario (i.e., it requires far less tuning and captures a substantially larger chunk of the phase space).

It is plausible that the decimation rules might allow for structures like limit cycles. It would also be interesting to see if other modes/states in the RSRG-X spectrum discussed present novel phases and to examine what typical properties of the system could be extracted through a study of states accessible via this scheme. 

It might also be worth looking into the case where we don't discard the measurement outcomes, and how the physics connects to that of monitored $U(1)$-symmetric random unitary circuits. Such circuits undergo charge sharpening transitions~\cite{sharpen} under $Z$-measurements. There has also been interest of late in strong to weak spontaneous symmetry breaking in such setups~\cite{strongweak2}. Such MIPTs are often transitions in the entanglement structure of measurement trajectories, and get washed out when looking at the density matrix alone. It must be noted however that in our problem, this sort of monitoring might lead to significantly different results on grounds that Hamiltonian evolution has more structure than random unitary evolution, and we measure $X$ and $Y$ rather than $Z$. Phases may also be considered in context of specific, individual quantum trajectories (with no averaging of any kind, including over entanglement entropy)-- say, the trajectory that returns the $+1$ eigenvalue at every measurement and working with weak measurements to allow for continuum limits. Writing out the quantum unraveling for this problem yields an effective Hamiltonian substantially different than the Lindbladian we studied, indicating different physics.

Under the Jordan-Wigner transformation, our model maps to a disordered free-fermionic chain, albeit with nonlocal measurements. In this light, it is worth mentioning that 1D ordered free-fermionic chains with local measurements do not exhibit phase transitions~\cite{freefermions1}, while disorder can stabilize criticality~\cite{freefermions2,discrit}. Also, nonlocal measurements do allow for novel phases in free fermionic systems~\cite{longrange,smallrange}.

Likewise, free fermions can still undergo MIPTs in higher dimensions~\cite{freefermions3}. Understanding in more detail the phases that exist under various kinds of measurements in higher dimensions is a worthwhile goal for theory. There has been past work on performing RSRG on higher dimensional lattices~\cite{highd1, highd2}. A similar numerical approach could be applied to our model by implementing our derived decimation rules on a 2D lattice, though an analytical treatment would likely be intractable because the lattice connectivity gets altered by decimation in higher dimensions. Also, the proliferation of long bonds would make the RG less controlled in more than one dimension, as for Hamiltonian systems~\cite{longbond}.

In closing, we turn to a brief discussion of possible experimental platforms for the realizations of measurement-induced dynamics with quenched disorder. Recent experiments have made headway in observing MIPTs in random unitary circuits on superconducting qubit arrays~\cite{exp2} and trapped-ion quantum computers~\cite{exp3}. Our model could potentially be trotterized to yield a circuit with quenched disorder, or directly realized, e.g. on a trapped ion system by pairwise coupling neighboring spins using lasers with random detunings to introduce disorder, along with stochastic fluorescence measurements. Since our dynamics averages over the measurement outcomes, we do not run into the post selection problem.

Random XX-type interactions can also be engineered using dipole-dipole exchange in Rydberg atom arrays and, post stochastic single-site measurements~\cite{exp6}, could provide an experimental avenue for testing the physics discussed here~\cite{exp1,exp5}. Measurements could be replaced by any form of engineered dephasing to add tunable phase noise competing with unitary evolution. Long-lived modes have also been of interest lately in the context of neutral atom chains~\cite{longlived1} and as an avenue for fast and high-fidelity quantum gates~\cite{longlived2}. We hope that the present work will stimulate further theoretical and experimental studies at the intersection of quenched disorder and quantum measurement.

\section*{Acknowledgements}
S.T. thanks Yi J. Zhao, Mytraya Gattu, Zack Weinstein, and Pushkar Mohile for discussions and comments. This work was supported by the NSF QLCI program through Grant No. OMA-2016245 to the Challenge Institute for Quantum Computation. The data that support the findings of this article are openly available~\cite{rsrg_x_scripts}.

\appendix
\section{Decimation Rules}
Here, we derive the decimation rules for generic configurations of cells. Specifically, we look at the case where some $J_i$ is the strongest coupling to be decimated, and when some $p_i/2$ is the strongest coupling.
\subsection{Decimating $J_i$}
The rules for when a type-$0$ cell is decimated were derived in the main text. Here, we look at the rules for decimating cell of types $1,2,3,$ and $4$.

For cells of types $1$, $2$, $3$, and $4$, respectively, the unperturbed bit is
\begin{multline}
\mathcal L_0^{(1)}=J_i [(\mathcal X_i x_{i+1}+\mathcal Y_i y_{i+1})+(x_i \mathcal X_{i+1}+y_i \mathcal Y_{i+1})\\-(\mathcal X_i \mathcal X_{i+1}+\mathcal Y_i \mathcal Y_{i+1})-(x_ix_{i+1}+y_iy_{i+1})],\\
\mathcal L_0^{(2)}=J_i [(\mathcal X_i x_{i+1}+\mathcal Y_i y_{i+1})+(x_i \mathcal X_{i+1}+y_i \mathcal Y_{i+1})\\+(\mathcal X_i \mathcal X_{i+1}+\mathcal Y_i \mathcal Y_{i+1})+(x_ix_{i+1}+y_iy_{i+1})],\\
\mathcal L_0^{(3)}=iJ_i [(\mathcal X_i x_{i+1}+\mathcal Y_i y_{i+1})-(x_i \mathcal X_{i+1}+y_i \mathcal Y_{i+1})\\+(\mathcal X_i \mathcal X_{i+1}+\mathcal Y_i \mathcal Y_{i+1})-(x_ix_{i+1}+y_iy_{i+1})],\\
\mathcal L_0^{(4)}=iJ_i [-(\mathcal X_i x_{i+1}+\mathcal Y_i y_{i+1})+(x_i \mathcal X_{i+1}+y_i \mathcal Y_{i+1})\\+(\mathcal X_i \mathcal X_{i+1}+\mathcal Y_i \mathcal Y_{i+1})-(x_ix_{i+1}+y_iy_{i+1})].
\end{multline}
On the other hand, exactly as before, the perturbation, 
\begin{multline}\mathcal L_\text{pert}=iJ_{i+1}[(\mathcal X_{i+1} \mathcal X_{i+2}+\mathcal Y_{i+1} \mathcal Y_{i+2})\\-(x_{i+1}x_{i+2}+y_{i+1}y_{i+2})]+iJ_{i-1}[(\mathcal X_{i-1} \mathcal X_{i}+\mathcal Y_{i-1} \mathcal Y_{i})\\-(x_{i-1}x_{i}+y_{i-1}y_{i})]+\frac{p_i}{2}\left(\mathcal X_i x_i+\mathcal Y_i y_i-2\mathbb I\right)\\+\frac{p_{i+1}}{2}\left(\mathcal X_{i+1} x_{i+1}+\mathcal Y_{i+1} y_{i+1}-2\mathbb I\right)+\dots,\end{multline}
with the dots standing in for terms that act trivially on the eigenbasis of $\mathcal L_0$.

$\mathcal L_0^{(3,4)}$ are imaginary, so we project to their ground states (lowest eigenvalues), $\ket{\text{GND}^{(3,4)}}$. Meanwhile, $\mathcal L_0^{(1,2)}$ have real eigenvalues, so we project to the longest lived state (highest eigenvalues), $\ket{\text{LL}^{(1,2)}}$. All of these states have the form of entangled modes $\pm\ket{00}\bra{00}\pm\sqrt{2}\ket{01}\bra{10}\pm\ket{01}\bra{01}\pm\ket{10}\bra{10}\pm\sqrt{2}\ket{10}\bra{01}\pm\ket{11}\bra{11}$ in the original Hilbert space.

If $\ket{m^{(1,2,3,4)}}$ is a general eigenstate of $\mathcal L_0^{(1,2,3,4)}$, use second order perturbation theory to write, post-decimation, \begin{multline}
    \mathcal L'=\bra{\text{GND/LL}}\mathcal L_0+\mathcal L_\text{pert}\ket{\text{GND/LL}}\\+\sum_{m\neq \text{GND/LL}}\frac{\bra{\text{GND/LL}}\mathcal L_\text{pert}\ket{m}\bra{m}\mathcal L_\text{pert}\ket{\text{GND/LL}}}{E_\text{GND/LL}-E_m}.
\end{multline}

The second order part of this expression yields the new coupling generated. The structure of the term tells us the cell-type post-decimation. The rest of the terms renormalize the ground state energy and the net dissipation rate.
\label{a1}
\subsection{Decimating $p_i/2$}
Here, \begin{equation}\mathcal L_0^{\text{vert}}=\frac{p_i}{2} \left[(\mathcal X_i x_i+\mathcal Y_i y_i)-2\mathbb I\right],\end{equation}
while
    
\begin{multline}\mathcal L_\text{pert}^{\text{vert}}=G_i(\mathcal X_i \mathcal X_{i+1}+\mathcal Y_i \mathcal Y_{i+1})+G_i^* (x_i x_{i+1}+y_i y_{i+1})\\+H_i(\mathcal X_i x_{i+1}+\mathcal Y_i y_{i+1})+H_i^* (x_i \mathcal X_{i+1}+y_i \mathcal Y_{i+1})\\+G_{i-1}(\mathcal{X}_i \mathcal{X}_{i-1} + \mathcal{Y}_i \mathcal{Y}_{i-1}) + G_{i-1}^* (x_i x_{i-1} + y_i y_{i-1}) \\+ H_{i-1}(\mathcal{X}_{i-1} x_i + \mathcal{Y}_{i-1} y_i) + H_{i-1}^* (\mathcal{X}_i x_{i-1} + \mathcal{Y}_i y_{i-1})
,\end{multline}
with $G_i,H_i\in\{\pm J_i, \pm i J_i\}$, depending on the types of the cells adjacent to the vertical bond. The star $(*)$ indicates complex conjugation.

Then, projecting to $\mathcal L_0^\text{vert}$'s longest lived (highest eigenvalue) eigenstate, which physically happens to be the maximally mixed state, we find an effective $\mathcal L'$ using second order perturbation theory as before, and extract the couplings obtained upon decimating a vertical bond.
\end{document}